\documentclass[a4paper,11pt]{article}
\pdfoutput=1
\usepackage{jheppub}
\usepackage[T1]{fontenc}
\usepackage[english]{babel}
\usepackage{amsthm,amsfonts}
\hyphenchar\font=\string"7F
\usepackage{breqn}
\usepackage{slashed}
\usepackage{stmaryrd}
\usepackage{enumerate}
\usepackage[normalem]{ulem}
\usepackage{cancel}
\usepackage{cite}
\usepackage{tikz}
\usetikzlibrary{arrows,matrix,calc}
\tikzset{>=stealth'}
\tikzstyle{shiftup}=[transform canvas={yshift=0.25em}]
\tikzstyle{shiftdn}=[transform canvas={yshift=-0.25em}]
\tikzstyle point=[minimum size=1mm,inner sep=0pt,outer sep=0pt,shape=circle,fill=black]

\usepackage{bbm} 
\usepackage{mathtools}
\usepackage{booktabs}
\usepackage{todonotes}

\DeclareMathOperator{\ad}{ad}
\DeclareMathOperator{\Tr}{Tr}

\newcommand{\DFTwzw}{DFT${}_\mathrm{WZW}$}
\allowdisplaybreaks
\newcommand{\HIDDEN}[1]{}

\newcommand{\qandq}{\qquad\mathrm{and}\qquad}

\newcommand{\dd}{\mathrm{d}}		
\newcommand{\DD}{\mathcal{D}}	
\newcommand{\PS}{\mathcal{P}}		
\newcommand{\HH}{\mathcal{H}}	
\newcommand{\NN}{\mathcal{N}}

\newcommand{\LL}{\mathcal{L}}	
\newcommand{\genlie}{\mathbb{L}}	
\newcommand{\Lt}{\tilde{L}}

\newcommand{\Pt}{\tilde{P}}

\newcommand{\id}{\mathbbm{1}}
\newcommand{\T}{\mathsf{T}}
\newcommand{\PdcV}{\Pi_{\mathtt{v}}}  %projectors for dressing coset
\newcommand{\PdcH}{\Pi_{\mathtt{h}}}

\newcommand{\BW}{B_{\mathrm{wzw}}}	
\newcommand{\HW}{H_{\mathrm{wzw}}}	
	
\newcommand{\bl}{[\![}
\newcommand{\br}{]\!]}
\newcommand{\bracd}{\bl\ ,\ \br}
\newcommand{\brac}{[\ ,\ ]}
\newcommand{\cwedge}{\overset{\wedge}{,}}

\title{\boldmath Para-Hermitian Geometries for \\ Poisson-Lie Symmetric $\sigma$-models}
\preprint{LMU-ASC 20/19\\MPP-2019-87}

\author[a]{Falk Hassler,}
\author[b,c]{Dieter L\"ust}
\author[b]{and Felix J. Rudolph}

\emailAdd{falk@fhassler.de}
\emailAdd{dieter.luest@lmu.de}
\emailAdd{frudolph@mpp.mpg.de}

\affiliation[a]{Department of Physics, University of Oviedo, Oviedo, E-33007, Spain}
\affiliation[b]{Max-Planck-Institut f\"ur Physik\\
F\"ohringer Ring 6, 80805 M\"unchen, Germany}
\affiliation[c]{Arnold-Sommerfeld-Center f\"ur Theoretische Physik\\
Department f\"ur Physik, Ludwig-Maximilians-Universit\"at M\"unchen\\
Theresienstra\ss e 37, 80333 M\"unchen, Germany}

\abstract{The doubled target space of the fundamental closed string is identified with its phase space and described by an almost para-Hermitian geometry. We explore this setup in the context of group manifolds which admit a maximally isotropic subgroup. This leads to a formulation of the Poisson-Lie $\sigma$-model and Poisson-Lie T-duality in terms of para-Hermitian geometry. The emphasis is put on so called half-integrable setups where only one of the Lagrangian subspaces of the doubled space has to be integrable. Using the dressing coset construction in Poisson-Lie T-duality, we extend our construction to more general coset spaces. This allows to explicitly obtain a huge class of para-Hermitian geometries. Each of them is automatically equipped which a generalized frame field, required for consistent generalized Scherk-Schwarz reductions. As examples we present integrable $\lambda$- and $\eta$-deformations on the three- and two-sphere.}

\begin{document}
\maketitle

\section{Introduction}
Physics and geometry are connected in an intriguing way. Perhaps the most prominent example is the intimate relation between general relativity and Riemannian geometry. There is a large variety of other examples ranging from gauge theories to condensed matter systems. In this paper we want to provide additional evidence for this paradigm by presenting a link between para-Hermitian geometry and Poisson-Lie $\sigma$-models. The latter were first studied because they admit Poisson-Lie T-duality \cite{Klimcik:1995ux,Klimcik:1995dy}, a generalization of abelian T-duality. 

Since this duality is not as well known as its abelian counterpart and comes with some additional subtleties, let us start by explaining its historical origins. Gauging the $\sigma$-model of a closed string moving in a target space with abelian isometries gives rise to a dual $\sigma$-model after applying a procedure due to Buscher \cite{Buscher:1987sk,Buscher:1987qj}. Classically, and even after quantization, the dynamics of both models is indistinguishable. This is remarkable because the dual target space looks in general quite different compared to the original one. Abelian T-duality has become an indispensable tool in studying string theory and therefore it is desirable to look for generalizations. The original Buscher procedure relies on abelian isometries, but in general isometry groups are non-abelian. Thus, extending it to non-abelian isometries results in in non-abelian T-duality \cite{delaOssa:1992vci}. There are however some additional challenges one has to cope with in this approach \cite{Giveon:1993ai,Alvarez:1994np,Elitzur:1994ri}. Most striking is that the dual target space lacks some of the isometries which would be required to go back to the original one \cite{Alvarez:1994dn}. Hence, non-abelian T-duality looks asymmetric and not symmetric like a duality should.

Poisson-Lie T-duality was introduced in \cite{Klimcik:1995ux,Klimcik:1995dy} to solve this problem. It embeds the physical target space into a higher dimensional Drinfeld double (a Lie group with special properties) and describes dual $\sigma$-models as different consistent embeddings. This way it considerably extends the class of target spaces which can be related by duality transformations. Poisson-Lie T-duality includes abelian and non-abelian T-duality as special cases but also goes beyond it. In particular it allows to connect certain target spaces which lack any isometries. However, it still only applies to a very restricted class of geometries which are called Poisson-Lie symmetric \cite{Klimcik:1995jn}. There is a crucial difference between between abelian and Poisson-Lie T-duality though. While the former is a genuinely symmetry of string theory and holds to all orders in $\alpha'$ and $g_s$ \cite{Rocek:1991ps}, the latter is in general restricted to the classical regime. This problem already applies for non-abelian T-duality \cite{Giveon:1993ai}. Whether it is possible to overcome it and eventually include at least some stringy corrections is an open question. However, Poisson-Lie T-duality preserve conformal invariance at the one-loop level \cite{Sfetsos:1998kr,Sfetsos:2009dj,Valent:2009nv} after imposing a mild unimodularity constraint and is therefore a powerful solution generating technique in supergravity\footnote{The non-unimodular case is governed by the generalized supergravity equations of motion \cite{Arutyunov:2015mqj}. While for making contact with full string theory the distinction between unimodular and non-unimodular is important, it is not relevant for the results presented in this paper.}. Applications include using non-abelian T-duality to generates new examples of holographic backgrounds \cite{Sfetsos:2010uq,Lozano:2011kb,Lozano:2012au,Lozano:2013oma,Itsios:2013wd}.

More recently, Poisson-Lie symmetry gained a lot of interest after its connection to integrable two-dimensional $\sigma$-models was discovered. This development started with the pioneering work by Klim\v{c}\'ik on the Yang-Baxter $\sigma$-model \cite{Klimcik:2002zj,Klimcik:2008eq} and attracted more attention after it was generalized to symmetric spaces and applied successfully to AdS$_5 \times$S$^5$\cite{Delduc:2013qra}. The undeformed version of this background represents the standard example for another important duality in string theory: the AdS/CFT correspondence \cite{Maldacena:1997re}. It relates closed strings in a $D$-dimensional anti de Sitter (AdS) spacetime with a conformal field theory (CFT) in $D$-$1$ dimensions. Studying both sides of this duality simultaneously is hard because one side is always strongly coupled and not accessible with perturbative techniques. However because AdS$_5 \times$S$^5$ is integrable \cite{Bena:2003wd}, it is still possible to make further progress in exploring the underlying principles of the AdS/CFT correspondence. A comprehensive review of this beautiful topic is given in \cite{Beisert:2010jr}. Considering this success, a natural question is if there are any other integrable $\sigma$-models for holographic backgrounds. This question is much harder to answer than one might initially think and triggered a lot activity recently. The standard approach is to start with one of the few known integrable models and deform them such that their classical integrability is preserved \cite{Delduc:2013fga}. The resulting integrable deformations fall into two distinct classes: the $\eta$-deformation \cite{Klimcik:2002zj} deforms a principal chiral model and the $\lambda$-deformation \cite{Sfetsos:2013wia} originates from a Wess-Zumino-Witten (WZW) model. Subsequently, both where shown to be connected to each other by applying Poisson-Lie T-duality and an analytic continuation \cite{Hoare:2015gda,Sfetsos:2015nya,Klimcik:2015gba}. Based on them, several multi-parameter deformations were introduced, for example the bi-Yang-Baxter model \cite{Klimcik:2014bta,Klimcik:2016rov} or the Yang-Baxter Wess-Zumino (YB WZ) model \cite{Kawaguchi:2011mz,Kawaguchi:2013gma,Delduc:2014uaa,Orlando:2016qqu,Klimcik:2017ken,Demulder:2017zhz}. All of them are captured by a Poisson-Lie $\sigma$-model.

There are different hints that these physical systems should have a natural relation to almost para-Hermitian geometry. Roughly speaking, para-Hermitian geometry is the real version of the more familiar concepts of complex, Hermitian and K\"ahler geometry. An almost para-complex structure $K$ on a $2d$-dimensional manifold is an endomorphism of the tangent bundle which squares to plus one, $K^2 = +\id$. It splits the tangent bundle into two eigenbundles of equal rank $d$. When in addition it is compatible with a metric $\eta$ of signature $(d,d)$, we have an almost para-Hermitian manifold. Together $K$ and $\eta$ give rise to the fundamental two-form $\omega$. If it is closed, $\dd\omega = 0$, we have an almost para-K\"ahler manifold. When the almost para-complex structure $K$ is integrable (i.e. its Nijenhuis tensor vanishes), we can drop the ``almost'' and are dealing with para-Hermitian or para-K\"ahler geometry.

The relation of this mathematical framework to physical systems can be seen in Double Field Theory (DFT) \cite{Siegel:1993xq,Siegel:1993th,Hull:2009mi,Hohm:2010pp,Hohm:2010xe}. This is a T-duality covariant effective target space theory of closed strings which requires a para-Hermitian structure or the slightly weaker half-integrable structure on the doubled space for consistency \cite{Vaisman:2012ke,Vaisman:2012px,Freidel:2017yuv,Freidel:2018tkj,Svoboda:2018rci}. A sketchy but short argument why this is the case goes as follows: recall that a complex structure on an even-dimensional real manifold allows us to introduce holomorphic and anti-holomorphic coordinates on this manifold. A para-Hermitian structure has a similar property, it allows for a splitting of the coordinates into what we will call physical and unphysical coordinates. In DFT all fields and parameters of gauge transformations are just allowed to depend on the physical coordinates. Only this way a necessary constraint (the section condition) is solved and the theory consistently reduces to normal supergravity. 

Historically DFT was derived in the context of abelian T-duality. This allows for the interpretation of its doubled space as being formed from directions which are conjugated to momentum and winding modes of the closed string \cite{Hull:2009mi}. But it can also be adapted so that it captures full Poisson-Lie T-duality \cite{Blumenhagen:2014gva,Blumenhagen:2015zma,Hassler:2017yza,Lust:2018jsx}. In this case the doubled space has rather the interpretation of a phase space for the underlying closed string theory. The same interpretation motivated metastring theory \cite{Freidel:2014qna,Freidel:2015pka} where para-Hermitian and Born geometry are crucial. By Born geometry one understands the addition of a Riemannian metric $\HH$ to a para-Hermitian manifold which is compatible with the para-Hermitian structures $(K,\eta,\omega)$. Thus it seems to be that para-Hermitian geometry and Poisson-Lie symmetry are closely related. Additional weight to this conjecture is given by \cite{Marotta:2018myj} where Drinfeld doubles, which play a central role in Poisson-Lie T-duality, are directly related to para-Hermitian geometry. 

We work out this connection in the present paper by constructing a para-Hermitian geometry with explicit expressions for $\eta$ and $\omega$ for every Poisson-Lie $\sigma$-model. Furthermore, we show that if the target space does not have $H$-flux this geometry becomes a Born geometry. The precise relation we establish between para-Hermitian and Born geometry, generalized geometry, DFT on group manifolds and the $\mathcal{E}$-model is of mutual benefit. From a physics point of view the underlying mathematical structure allows for a unified description of dual backgrounds. In general T-duality does not only connect two distinct target spaces but is a plurality identifying complete families of them. We will explain how deformation theory of the para-Hermitian structure provides a powerful tool to explore the resulting moduli space of dual backgrounds. On the other hand Poisson-Lie $\sigma$-models provide a rich class of examples for para-Hermitian geometries which can be explicitly constructed on group manifolds. We use them to prove that certain constraints imposed in the recent works on para-Hermitian structures and their connection to generalized geometry \cite{Hitchin:2004ut,Gualtieri:2003dx} and DFT \cite{Freidel:2017yuv,Freidel:2018tkj,Svoboda:2018rci,Marotta:2018myj} can be relaxed. In particular, we find that only the distribution $\Lt$ associated to the unphysical coordinates has to be integrable to permit the construction of a Courant algebroid over the physical target space. Finally, we study the doubled geometry of gauged Wess-Zumino-Witten models related to the dressing coset construction \cite{Klimcik:1996np,Squellari:2011dg,Klimcik:2019kkf}. The latter provides the most general class of target spaces with Poisson-Lie symmetry. For example the above mentioned deformations of AdS$_5 \times$S$^5$ arise from this construction. All para-Hermitian geometries we construct come automatically with a generalized frame field which can be used to find new, consistent generalized Scherk-Schwarz reductions \cite{Grana:2012rr,Geissbuhler:2011mx,Aldazabal:2011nj}. 

The paper is organized as follows: in section \ref{sec:GeometryRecap} we will briefly review the settings of (almost) para-Hermitian geometry and Born geometry including the important notion of a D-structure. In section \ref{sec:GroupManifolds} we will discuss how these objects arise very natural on certain group manifolds. Here we will comment on the relation to Poisson-Lie $\sigma$-models, generalized geometry and start looking at Poisson-Lie symmetry and its relevance for the integrability of a $\sigma$-model. From there we move on to dressing cosets and their doubled geometry in section~\ref{sec:dressingcoset}. Finally, we will explore how to deform these setups by keeping the required half-integrable structure in section~\ref{sec:Deformation} and present the YB WZ model as an explicit examples in section~\ref{sec:examples}. We conclude in the last section.

\section{Para-Hermitian and Born geometry}
\label{sec:GeometryRecap}
This section presents the mathematical background which we apply in the next section to the Poisson-Lie $\sigma$-model. We summarize the key ideas of para-Hermitian geometry together with the concept of a D-structure and suitably generalized notions of torsion and integrability\footnote{In this paper two notions of integrability are relevant. First there is integrability of a physical theory which requires an appropriate number of integrals of motion. Second there is also integrability for distributions in a mathematical sense. The latter is important for this section.}. This forms the basis to define Born geometry and has been established in \cite{Freidel:2017yuv,Svoboda:2018rci,Freidel:2018tkj}, for an executive summary see \cite{Svoboda:2019fpt}.

\subsection{Para-Hermitian geometry}
\label{sec:paraHermitianGeometry}
We start with an \emph{almost para-complex manifold} $(\PS, K)$ where $\PS$ is a $2d$-dimensional differentiable manifold and $K$ an endomorphism on the tangent bundle $T\PS$ such that $K^2=+\id$. The $+1$ and $-1$-eigenbundles of $K$ have the same rank and will be denoted by $L$ and $\Lt$ respectively. The associated projection operators are 
\begin{equation}\label{eqn:P&tildeP}
  P= \frac12(\id + K) \qandq \Pt = \frac12(\id - K)\,.
\end{equation}
The integrability of the para-complex structure $K$ can be expressed in terms of the Nijenhuis tensor analogous to the complex case
\begin{align}
\begin{aligned}
  N_K(X,Y)&\coloneqq\frac{1}{4}\Big( [X,Y]+[KX,KY]-K([KX,Y]+[X,KY])\Big)\\
    &=P[\Pt X,\Pt Y]+\Pt [P X,P Y]
\end{aligned}
\label{eq:nijenhuis}
\end{align}
where $\brac$ is the Lie bracket on $T\PS$. The para-complex structure $K$ is integrable if and only if $N_K$ vanishes. Then the eigenbundles are integrable distributions and we say $(\PS,K)$ is a para-complex manifold. 

An important difference from almost complex manifolds is that the integrability of $L$ is independent from the integrability of $\Lt$: $L$ can be integrable when $\Lt$ is not and vice versa. This can be seen clearly in the second line of the Nijenhuis tensor above where the two terms live in $L$ and $\Lt$ respectively and can vanish independently. This leads to the notion of half-integrability: we say an almost para-complex manifold $(\PS,K)$ is $L$-integrable ($\Lt$-integrable) if $L$ ($\Lt$) is an integrable distribution. This feature of half-integrability will be crucial later on. We will also see that some (T-duality) transformations such as a B-transformation may not preserve integrability and can thus change the integrability of the subspace. Furthermore, the appearance of fluxes can be related to obstruction to the integrability of $L$. 

Next we add a pseudo-Riemannian metric $\eta$ compatible with the almost para-complex structure $K$ to our setup to obtain an \emph{almost para-Hermitian} manifold $(\PS, \eta,K)$. The compatibility of the two structures is expressed by the skew-orthogonality of $K$ with respect to $\eta$
\begin{equation}
  \eta(KX,KY)=-\eta(X,Y)\,.
  \label{eq:etaKK}
\end{equation}
Due to this skewness of $K$ and the fact that its eigenbundles have the same rank, the metric $\eta$ is of split signature $(d,d)$. Together the two structures define the tensor $\omega\coloneqq \eta K$ which is skew and non-degenerate. Therefore $\omega$ is an almost symplectic form which is anti-compatible with $K$
\begin{equation}
  \omega(KX,KY)=-\omega(X,Y)\,.
  \label{eq:omegaKK}
\end{equation}
It follows that the eigenbundles $L$ and $\Lt$ are isotropic subspaces with respect to $\eta$ and $\omega$. They are null and Lagrangian respectively. 

\subsection{D-structure}\label{sec:Dstructure}
The next step is to consider a generalized differentiable structure for para-Hermitian geometry whose corresponding bracket operation is the analogue of the Lie bracket \cite{Vaisman:2004msa,Vaisman:2012ke,Vaisman:2012px,Freidel:2017yuv,Svoboda:2018rci}. To this end, we start with a \emph{metric-compatible bracket} defined on any pseudo-Riemannian manifold $(\PS,\eta)$. This is a bilinear operation $\bracd:\Gamma(T\PS)\times \Gamma(T\PS) \rightarrow \Gamma(T\PS)$ on the algebra of vector fields which is compatible with the metric and satisfies a normalization condition
\begin{align}
  X[\eta(Y,Z)] &= \eta(\bl X,Y\br,Z)+\eta(Y,\bl X,Z\br), \label{eq:DbracketComp} \\
  \bl X,X \br &= \frac{1}{2}\mathcal{D}[\eta(X,X)] \label{eq:DbracketNorm}
\end{align}
where we associate to any function $f$ a vector field $\DD[f]$ defined by $\eta(\DD[f],X)=X[f]$. Note that the bracket is not skew-symmetric so the Leibniz property takes the form	
\begin{align}
  \bl X,fY \br &= f\bl X,Y\br+X[f]Y, \label{eq:DbracketLeibniz1} \\
  \bl fX,Y\br &= f\bl X,Y\br-Y[f]X+\eta(X,Y)\mathcal{D}[f]\,.\label{eq:DbracketLeibniz2}
\end{align}
One can construct a metric-compatible bracket for any metric-compatible connection $\nabla$ (one which satisfies $\nabla_X\eta=0$) via
\begin{equation}
  \eta(\bl X,Y\br^\nabla,Z)=\eta(\nabla_XY-\nabla_YX,Z)+\eta(\nabla_ZX,Y)\,.
  \label{eq:D-bracket}
\end{equation}
This metric-compatible bracket can be used to define a generalized notion of integrability for any endomorphism $C$ on $T\PS$ which satisfies $C^2=\pm\id$ and $\eta(CX,Y) = -\eta(X,CY)$, like our almost para-complex structure $K$. The \emph{generalized Nijenhuis tensor} associated to such an object $C$ is defined by
\begin{equation}\label{eqn:genNijenhuis}
  \NN_C(X,Y) \coloneqq \frac14 \Big( C^2\bl X,Y\br +\bl C X,C Y\br  - C\big(\bl C X,Y\br  + \bl X,C Y\br\big) \Big)\,.
\end{equation}
One can show that this indeed defines a skew-symmetric tensor which is only the case for the above sign choice for the orthogonality of $C$ with respect to $\eta$ and is a non-trivial result since the bracket is not skew-symmetric itself (unlike for the usual Nijenhuis tensor \eqref{eq:nijenhuis} in terms of the Lie bracket) \cite{Freidel:2018tkj}.

With a notion of generalized integrability in place we can further refine our bracket. A \emph{D-bracket} on an almost para-Hermitian manifold $(\PS,\eta,K)$ is a metric-compatible bracket $\bracd$ such that $K$ is integrable in the generalized sense ($\NN_K=0$). When this is the case, the data $(\PS,\eta,K,\bracd)$ has been called a \emph{D-structure} \cite{Freidel:2018tkj}. If we have a D-structure, the subbundles $L$ and $\Lt$ are Dirac structures with respect to the D-bracket ($\bl L,L\br \subset L$ and $\bl \Lt,\Lt\br \subset \Lt$). In other words, they are individually integrable in the generalized sense 
\begin{equation}
  \begin{aligned}
    \NN_K(PX,PY) &= \Pt(\bl PX, PY\br) = 0, \\
    \NN_K(\Pt X,\Pt Y) &= P(\bl \Pt X, \Pt Y\br) = 0 \,.
  \end{aligned}
  \label{eq:GeneralisedIntegrability}
\end{equation}

If in addition to all the properties mentioned so far a D-bracket also satisfies
\begin{equation}
  \begin{aligned}
    \bl PX,PY \br&= P([PX,PY]),\\
    \bl \Pt X,\Pt Y \br&=\Pt([ \Pt X,\Pt Y])\,,
    \label{eq:canonical}
  \end{aligned}
\end{equation}
it is said to be \emph{canonical}. In this case when it is restricted to $L$ (respectively $\Lt$), it reduces to the projection of the Lie bracket onto $L$ (respectively $\Lt$). It turns out that there is a unique canonical D-bracket \cite{Freidel:2018tkj}. This canonical D-bracket is the D-bracket that appears in the DFT literature and is closely related to the Dorfman bracket of Courant algebroids in generalized geometry \cite{Dorfman, Severa:2017oew, Grana:2008yw, Hull:2009zb}. 

The D-bracket can also be seen as a \emph{generalized Lie derivative} acting on a vector which we will denote by $\genlie$
\begin{equation}\label{eqn:genLie}
  \bl X, Y \br = \genlie_XY \,.
\end{equation}
As for the Lie derivative, its action can easily be extended to arbitrary tensors by using the Leibniz property it inherits from the D-bracket.

\subsection{Born geometry}
\label{sec:BornGeometry}
The (almost) para-Hermitian geometry we have discussed so far together with a D-structure provides the kinematical structure of the doubled geometry. In a final step we now introduce a generalized metric $\HH$ which contains the dynamical degrees of freedom. Under particular circumstances which we explain now, the triple $(\eta,\omega,\HH)$ on the manifold $\PS$ forms what has been named a \emph{Born geometry}. Whereas in (almost) para-Hermitian geometry we could make some global statements and considered integrability conditions such as $N_K=0$, in Born geometry we will only consider $(\eta,\omega,\HH)$ on local patches.

The additional Riemannian metric $\HH$ has to be compatible with $\eta$ and $\omega$ by satisfying
\begin{equation}
  \eta^{-1}\HH=\HH^{-1}\eta \qandq \omega^{-1}\HH=-\HH^{-1}\omega .
\end{equation}
Note that here we view $(\eta,\omega,\HH)$ as maps $T\PS\to T^*\PS$. One can now define two more endomorphisms on the tangent bundle $T\PS$. Together with the almost symplectic form $\omega$, the generalized metric $\HH$ forms an almost Hermitian structure $(\omega,\HH, I)$
\begin{equation}
  \omega(IX,Y) = -\HH(X,Y), \qquad I^2 =-\id
\end{equation}
where $I$ is an almost complex structure. One also finds that the two metrics $\eta$ and $\HH$ form a chiral structure $J$ with
\begin{equation}
  \eta(JX,Y) = \HH(X,Y), \qquad J^2 =+\id \,.
  \label{eq:chiralstructure}
\end{equation}
The data $(\eta,\HH,J)$ is familiar from DFT. We call it a chiral structure since $J$ relates the left- and right-moving sector of the closed string. From section~\ref{sec:paraHermitianGeometry} above we already have the almost para-Hermitian structure $(\eta,\omega,K)$ with
\begin{equation}
  \omega(KX,Y) = \eta(X,Y), \qquad K^2 =+\id \,.
\end{equation}
The three endomorphisms $(I,J,K)$ on $T\PS$ all anti-commute with each other and additionally satisfy
\begin{equation}
  IJK = -\id
\end{equation}
so that they form an almost para-quaternionic structure. The key relations between the structures of Born geometry are summarized in table~\ref{tab:BornGeo}.
\begin{table}[!t]
\renewcommand{\arraystretch}{1.7}
\centering
\resizebox{0.95\textwidth}{!}{%
\begin{tabular}{@{}ccc@{}}
\toprule
$I={\HH}^{-1}{\omega}=-\omega^{-1}\HH$ & $J={\eta}^{-1}{\HH}={\HH}^{-1}{\eta}$ & $K={\eta}^{-1}{\omega}={\omega}^{-1}{\eta}$  \vspace{8pt} \\
$-I^2=J^2=K^2=\id$                     & $\{I,J\}=\{J,K\}=\{K,I\}=0$           & $IJK=-\id$     \vspace{8pt} \\
$\HH(IX,IY)=\HH(X,Y)$              & $\eta(IX,IY)= -\eta(X,Y)$         & $\omega(IX,IY)=\omega(X,Y)$             \\
$\HH(JX,JY)=\HH(X,Y)$              & $\eta(JX,JY)= \eta(X,Y)$          & $\omega(JX,JY)=-\omega(X,Y)$            \\
$\HH(KX,KY)=\HH(X,Y)$              & $\eta(KX,KY)= -\eta(X,Y)$         & $\omega(KX,KY)=-\omega(X,Y)$            \\ \bottomrule
\end{tabular}%
}
\caption{Summary of structures in Born geometry. Here $\{\ , \ \}$ is the anti-commutator.}
\label{tab:BornGeo}
\end{table}%
Another interesting and important property of Born geometry is the following. The triple $(\eta,\omega,\HH)$ is a Born structure on $\PS$ if and only if there exists a frame $E\in GL(2d)$ such that one can write $\eta = E^\T \bar{\eta} E$,  $\omega =E^\T\bar\omega E$ and $\HH=E^\T\bar{\HH} E$ with 
\begin{align}
  \bar\eta &= 
\begin{pmatrix}
	0&\id\\
	\id&0
\end{pmatrix} \, , &
\bar\omega &=
\begin{pmatrix}
	0&-\id\\
	\id&0
\end{pmatrix} \, , &
\bar\HH =
\left(
\begin{array}{cc}
	\id &0\\
	0& \id
\end{array}
\right) \, , \label{m}
\end{align}
in the case of Euclidean signature \cite{Freidel:2018tkj}. $\bar\eta$ and $\bar\omega$ are important constituents of the action for $\sigma$-model on a doubled target space \cite{Tseytlin:1990nb,Tseytlin:1990va,Giveon:1991jj,Hull:2006va}. We come back to this point in section~\ref{sec:PLsigmamodel}.

\section{Group manifolds and para-Hermitian geometry}
\label{sec:GroupManifolds}
We now show that certain group manifolds provide explicit examples of the structures introduced in the previous section. Furthermore, we relate them to $\sigma$-models whose target space admit Poisson-Lie T-duality \cite{Klimcik:1995ux,Klimcik:1995dy}. There is an intriguing interplay between the mathematical structure in para-Hermitian geometry and the physical properties of the $\sigma$-models. For example we will show that the D-bracket captures their global symmetries and that the generalized metric encodes the metric and two-form field on their target space. Finally, we discuss the conditions on a target space to permit Poisson-Lie T-duality in which case it is said to be Poisson-Lie symmetric \cite{Klimcik:1995jn}. It is in general hard to check for this symmetry directly at the target space level. But if we apply the framework of para-Hermitian geometry, we find that Poisson-Lie symmetry is equivalent to isometries of the generalized metric $\mathcal{H}$.

\subsection{\texorpdfstring{$\Lt$}{Ltilde}-integrable para-Hermitian Lie groups}\label{sec:parahermliegroup}
We start with a $2 d$-dimensional Lie group $\PS$ with group elements $p$. Instead of working directly with the tangent bundle of $\PS$, we use the globally defined left-invariant Maurer-Cartan form $E$, 
\begin{equation}
  T_A E^A (X) = p^{-1} \dd p \, (X) \,.
  \label{eq:MCform}
\end{equation}
The group $\PS$ is generated by the Lie algebra $\mathfrak{p}$ with the $2 d$ generators $T_A$. They satisfy the commutation relation
\begin{equation}
  [ T_A, T_B ] = F_{AB}{}^C T_C
\end{equation}
and all the relevant local data is encoded in the structure coefficients $F_{AB}{}^C$. The Maurer-Cartan form $E$ is a bundle isomorphism $E: T\PS \rightarrow \ad \PS$ whose inverse is $E^{-1}$. To make contact with a para-Hermitian structure, we need to define the projectors $P$ and $\Pt$ introduced in \eqref{eqn:P&tildeP} which project onto the subspaces $L$ and $\Lt$ of $T\PS$. Here $\Pt$ is chosen such that the image of $E \Pt$ is restricted to a $d$-dimensional subalgebra $\mathfrak{l}$ of $\mathfrak{p}$.

The pseudo-Riemannian metric $\eta$ is now given in terms of an ad-invariant, non-degenerate, symmetric pairing $\langle \,,\, \rangle$ on $\mathfrak{p}$ for which $\mathfrak{l}$ is maximally isotropic:
\begin{equation}
  \eta(X, Y) = \langle E X, E Y \rangle \,. \label{eqn:etaXY}
\end{equation}
To split the coordinates of $\PS$ in a physical and an unphysical part, we decompose the group element $p$ according to
\begin{equation}\label{eqn:paramP}
  p = \ell m \qquad \ell \in \mathcal{L}\,,\quad m \in M = \mathcal{L} \backslash \PS
\end{equation}
where $\ell$ is an element of the maximally isotropic subgroup $\LL$ and $m$ a coset representative. In terms of this splitting we define the symplectic form $\omega$ on $\PS$ as
\begin{equation}
  \omega(X,Y)  =   \langle \dd m m^{-1} (X) \cwedge \ell^{-1} \dd \ell (Y) \rangle +  \BW (X,Y) 
  \label{eqn:omegaXY} \,,
\end{equation}
where the 2-form $\BW$ is chosen such that locally
\begin{equation}\label{eqn:HW}
  \dd \BW = \frac16 \langle [ \dd m m^{-1}, \dd m m^{-1} ], \dd m m^{-1} \rangle = \HW
\end{equation}
holds. The notation $\langle\ \cwedge\, \rangle$ denotes the antisymmetrization of the pairing $\langle\ ,\, \rangle$. Our choice for $\eta$ and $\omega$ might seem a bit artificial, but we will see in the next subsection that it is well motivated when studying the Poisson-Lie $\sigma$-model.

First though, we have to verify that we indeed describe a half-integral para-Hermitian structure. There are two constraints. We have to calculate $K=\eta^{-1}\omega$ and check that it is an involution. Furthermore, the Nijenhuis tensor has to vanish, at least for one of the distributions $L$ or $\Lt$. The left action of $\LL$ on $\PS$ is transitive. Thus $M$ is a homogeneous space and we understand $\PS$ as a $\LL$-principal bundle over $M$. In each patch we use the local trivialization to split the coordinates $X^I = ( x^i \,\, \tilde x^{\tilde i} )$ on $\PS$ into a base contribution $x^i$ (physical) and a fiber part $\tilde x^{i}$ (unphysical). Expressing $\omega$ and $\eta$ explicitly in terms of these coordinates gives rise to
\begin{equation}
  \eta_{IJ} = \begin{pmatrix} \eta_{ij} & \eta_{i\tilde j} \\
    \eta_{\tilde i j} & 0 \end{pmatrix}\,, \quad
  \omega_{IJ} = \begin{pmatrix} \BW{}_{ij} &  \eta_{i\tilde j} \\
   - \eta_{\tilde i j} & 0 \end{pmatrix} \quad \text{and} \quad
  K^I{}_J = \begin{pmatrix} - \delta^i_j & 0 \\
    \eta^{\tilde i k}(\BW{}_{kj} + \eta_{k j}) & \delta^{\tilde i}_{\tilde j}
  \end{pmatrix}
\end{equation}
with
\begin{equation}
  \eta_{ij} = \langle \partial_i m m^{-1} , \partial_j m m^{-1} \rangle \,, \quad
  \eta_{i\tilde j} = \langle \partial_i m m^{-1} , \ell^{-1} \partial_{\tilde j} \ell \rangle  \quad \text{where} \quad
  \eta_{i\tilde k} \eta^{j\tilde k} = \delta_i^j\,.
\end{equation}
It is now straightforward to verify $K^2=\id$ or equally the compatibility of $\eta$ and $\omega$ with each other (and $K$) in \eqref{eq:etaKK} and \eqref{eq:omegaKK}. One is able to calculates the Nijenhuis tensors \eqref{eq:nijenhuis} and observe that its only non-vanishing contribution is
\begin{equation}
  (N_K)^{\tilde i}{}_{j k} = 2 K^{l}{}_{[j} \partial_{l} K^{\tilde i}{}_{k]} + 2 K^{\tilde l}{}_{[j} \partial_{\tilde l} K^{\tilde i}{}_{k]}
 										 - 2 K^{\tilde i}{}_{\tilde l} \partial_{[j} K^{\tilde l}{}_{k]}\,.
\end{equation}
In particular, the Nijenhuis tensor is annihilated by $\tilde P$. As a consequence $ P [ \Pt X, \Pt Y ] = 0$ holds, while in general $\Pt [ P X, P Y ] = 0$ is violated. In this case we are dealing with a half-integrable structure. Half-integrable structures were studied recently \cite{Freidel:2018tkj,Marotta:2018myj}. But in those references, integrability with respect to $L$ instead of $\Lt$ is required. Here we find the opposite situation. In the next section we will see that there is still no problem in making contact with the physics of Poisson-Lie $\sigma$-models. Eventually, we are even able to construct a canonical Courant algebroid just based on $\Lt$-integrability.

There are two special cases where we can choose a coset representative $m$ such that
\begin{equation}\label{eqn:addrestriction}
  \eta_{ij} =  0
\end{equation}
holds:
\begin{itemize}
  \item \emph{Drinfeld doubles} are Lie groups with two maximally isotropic subgroups. In this case, $\PS/\LL$ is a Lie group and $m$ is just an element of this Lie group.
  \item \emph{Pseudo Riemannian symmetric spaces} arise for $\PS/\LL$ if the Lie algebra $\mathfrak{l}$ and its complement $\mathfrak{m}$ form a symmetric pair
    \begin{equation}
      [\mathfrak{m}, \mathfrak{m}] = \mathfrak{l} \qquad
      [\mathfrak{l}, \mathfrak{m}] = \mathfrak{m} \qquad
      [\mathfrak{l}, \mathfrak{l}] = \mathfrak{l}\,.
    \end{equation}
    In this case, $m$ arises from applying the exponential map to $\mathfrak{m}$:
    \begin{equation}
      m = \exp(\mathfrak{m})\,.
    \end{equation}
\end{itemize}
Only in the first case $\BW$ vanishes as well as $\eta_{ij}$. In this case, which was also studied in \cite{Marotta:2018myj}, both $L$ and $\Lt$ are integrable and we obtain a para-Hermitian structure. We conclude that the obstruction of $L$ to be integrable is measure by $\HW$ in \eqref{eqn:HW}. As we show in the next subsection, this three-form represents the WZ-term of the worldsheet two-dimensional $\sigma$-model. It is classified by the third de~Rham cohomology $H^3(\PS/\LL)$.

\subsection{Poisson-Lie \texorpdfstring{$\sigma$}{sigma}-model}\label{sec:PLsigmamodel}
Intriguingly, the structures presented in the last subsection appear also in a class of string theory worldsheet models. They are called Poisson-Lie $\sigma$-models and were introduced by Klim\v{c}\'ik and Severa in \cite{Klimcik:1995ux,Klimcik:1995dy,Klimcik:1996nq}. Their dynamics is governed by the action
\begin{equation}\label{eqn:Ssigma}
  S = \frac12 \int \dd\sigma\,\dd\tau \Big( \eta( X_\sigma, X_\tau ) + \omega( X_\sigma, X_\tau ) - \HH( X_\sigma, X_\sigma ) \Big)
\end{equation}
with
\begin{equation}\label{eqn:XsigmaXtau}
  X_\sigma = E^{-1} ( p^{-1} \partial_\sigma p )  \quad \text{and} \quad
  X_\tau = E^{-1} ( p^{-1} \partial_\tau p )\,,
\end{equation}
and $\HH$ denoting the generalized metric introduced in section~\ref{sec:BornGeometry}. An action of this form is typical for a $\sigma$-model on a doubled target space \cite{Tseytlin:1990nb,Tseytlin:1990va} (the topological term was later considered by  \cite{Giveon:1991jj,Hull:2006va}).

There is an alternative way of writing this action in terms of a WZW-model on $\PS$ combined with an additional contribution from an involution $\mathcal{E}$. The latter captures the geometry of the target space and is in one-to-one correspondence with the generalized metric
\begin{equation}\label{eqn:EtoH}
  \HH(X,Y)=\langle E X, \mathcal{E} E Y \rangle\,.
\end{equation}
It plays the same role as the chiral structure $J$ in Born geometry (see \eqref{eq:chiralstructure}). The corresponding $\sigma$-model goes also under the name $\mathcal{E}$-model \cite{Klimcik:1995dy,Klimcik:1996nq,Klimcik:2015gba}. It is either captured by the action
\begin{equation}\label{eqn:SEmodel}
  S = \frac12 \int_\Sigma \dd\sigma\,\dd\tau \langle p^{-1} \partial_\sigma p, p^{-1} \partial_\tau p \rangle + \frac1{12} \int_{M_3} \langle [ p^{-1} \dd p , p^{-1} \dd p ], p^{-1} \dd p  \rangle - 
    \int \dd \tau \mathrm{Ham}
\end{equation}
or alternatively by the Hamiltonian
\begin{equation}
  \mathrm{Ham} = \frac12 \oint \dd\sigma \langle j(\sigma) , \mathcal{E} j(\sigma) \rangle
\end{equation}
where the currents $j(\sigma) = p^{-1} \partial_\sigma p$ are governed by the Poisson-bracket
\begin{equation}
  \{ j_A(\sigma), j_B(\sigma') \} = F_{AB}{}^C j_C(\sigma) \delta( \sigma - \sigma' ) + \eta_{AB} \delta'( \sigma - \sigma' )
\end{equation}
with $j_A(\sigma)$=$\langle T_A , j(\sigma)\rangle$ and $\eta_{AB} = \langle T_A, T_B \rangle$. The action involves a WZ-term which is evaluated on a three dimensional extension $M_3$ of the world sheet $\Sigma$ ($\partial M_3$=$\Sigma$)\footnote{This extension is in general not unique. Different extensions are labeled by elements of $\pi_3(\PS)$, the third homotopy group of $\PS$ \cite{Witten:1983tw}. The actions S(X) and S'(X) for different embeddings just differ by a constant. Thus the classical dynamics of the closed string does not depend on the particular choice of the extension.}

In order to identify the two actions \eqref{eqn:Ssigma} and \eqref{eqn:SEmodel}, we introduce the closed three-form
\begin{equation}
  F(X,Y,Z) = \langle [ E X, E Y ], E Z \rangle = 
    \frac16 \langle [ p^{-1} \dd p(X) , p^{-1} \dd p(Y) ], p^{-1} \dd p(Z)  \rangle 
  \label{eq:F}
\end{equation}
and notice that exterior derivative of $\omega$ gives rise to
\begin{equation}\label{eqn:domega}
  \dd \omega = F \,.
\end{equation}
In deriving this relation, we took into account that $\ell$ is an element of a maximally isotropic subgroup. It allows us to write locally the WZ-term in \eqref{eqn:Ssigma} as
\begin{equation}
  \frac12 \int_{M_3} X^* F = \frac12 \int_\Sigma X^* \omega = \frac12 \int_\Sigma \dd \sigma \, \dd \tau \, \omega( X_\sigma, X_\tau )\,.
\end{equation}
Here $X^*\omega$ is the pullback of $\omega$ to the worldsheet with $X$ being the usual embedding map into the target space. The crucial property of the $\mathcal{E}$-model is that not all its degrees of freedom are dynamical. Against the first intuition, it does not describe strings propagating in the target space $\PS$, but only in the coset $M$=$\mathcal{L}\backslash\PS$. For this reduction to take place, $\PS$ has to admit an almost para-Hermitian structure. Let us take a closer look at the sum $\eta(X_\sigma, X_\tau) + \omega(X_\sigma, X_\tau)$ to see how this works. If we use the splitting into coset and subgroup part in \eqref{eqn:paramP}, this sum expands to
\begin{equation}
  \underbrace{\langle \partial_\sigma m m^{-1} + \Lambda, \partial_\tau m m^{-1} \rangle + 
    \cancel{\langle \partial_\sigma m m^{-1} , \ell^{-1} \partial_\tau \ell \rangle}}_{\displaystyle \eta( X_\sigma, X_\tau )}
  + 
  \underbrace{\langle \Lambda, \partial_\tau m m^{-1} \rangle
  - \cancel{\langle \partial_\sigma m m^{-1}, \ell^{-1} \partial_\tau \ell \rangle}}_{\displaystyle \omega( X_\sigma, X_\tau )}
\end{equation}
with $\Lambda = \ell^{-1} \partial_\sigma \ell$. The two terms which cancel are the only ones in the action with a $\tau$ derivative action on $\ell$. Since there are no additional derivatives acting on $\Lambda$ it can be integrated out from the action \cite{Klimcik:1996nq} to obtain a $\sigma$-model with target space $M$.

$\mathcal{E}$-models make Poisson-Lie symmetry and T-duality manifest at the level of the worldsheet theory. They also admit to describe integrable deformations of principal chiral models and WZW models in a unified way. Before we discuss these applications, we relate their global symmetries to the unique D-structure of section~\ref{sec:Dstructure}.

\subsection{D-structure}\label{sec:Dstructure2}
A para-Hermitian structure on group manifolds admits a unique D-structure which can be constructed in the following way. First we fix an $\eta$-compatible connection
\begin{equation}
  \eta(\nabla_X Y, Z) = \eta(D_X Y, Z) - \frac13 F(X,Y,Z)\,.
\end{equation}
There are many other $\eta$-compatible connections on $\PS$. Why do we choose this one? The short answer is because it eventually gives rise to the unique D-structure described in section~\ref{sec:Dstructure}. It is also motivated by DFT on group manifolds \cite{Blumenhagen:2014gva} where this particular connection appeared for the first time. More recently it was described in the context of DFT in the supermanifold formulation \cite{Crow-Watamura:2018liw}. While this connection has curvature and torsion, $D$ is an $\eta$-compatible, flat connection on $\PS$. Thus, its curvature vanishes and its torsion given by
\begin{equation}
\eta(T_D(X,Y),Z) = \eta(D_XY-D_YX-[X,Y],Z) = F(X,Y,Z) \,,
\end{equation}
where $F$ is the three-form in \eqref{eq:F} which captures the structure constants of $\PS$. Using \eqref{eq:D-bracket}, the resulting metric-compatible bracket reads
\begin{equation}
  \eta(\bl X, Y \br^\nabla, Z) = \eta ( [X, Y], Z ) + \eta ( D_Z X , Y )\,
  \label{eq:DbracketonG}
\end{equation}
with $[X,Y]$ denoting the ordinary Lie bracket on $T\PS$.

These definitions permit to calculate the generalized Nijenhuis tensor \eqref{eqn:genNijenhuis}. Taking into account that it is skew-symmetric, its two contributions simplify to
\begin{align}
  \eta(\Pt \bl PX, PY \br , Z) &= \Pt N_K ( X, Y ) + \frac12 F( P X, P Y, P Z ) \label{eqn:PtgenNij}\\
  \intertext{and}
  \eta(P \bl \Pt X,\Pt Y \br, Z) &= P N_K ( X, Y ) + \frac12 F( \Pt X, \Pt Y, \Pt Z) = 0 \,.
\end{align}
The term $P N_K(X,Y)$ vanishes since $K$ is $\tilde L$-integrable in the ordinary sense (cf. \eqref{eq:nijenhuis}). At the same time 
\begin{equation}
  F(\Pt X, \Pt Y, \Pt Z) = 0
\end{equation}
does not contribute because we require $\mathcal{L}$ to be a maximally isotropic subgroup of $\PS$. However the second part \eqref{eqn:PtgenNij} of the generalized Nijenhuis tensor does not vanish in general. Therefore, the D-structure is only $\tilde L$-integrable. A notable exception are Drinfeld doubles where $\mathcal{L}$ and $\mathcal{L}\backslash\PS$ are both maximally isotropic subgroups. Their $H$-flux is trivial in de Rham cohomology and their $\sigma$-model does not posses a WZ-term. The general lack of $L$-integrability does not pose a problem because -- as we will show next -- the D-bracket \eqref{eq:DbracketonG} still reduces to the canonical Dorfman bracket on generalized tangent space $T M\oplus T^* M$ of the target space $M$=$\mathcal{L}\backslash\PS$.

\subsubsection*{Generalized geometry on $T M\oplus T^* M$}\label{sec:gengeometry}
In order to connect the D-bracket \eqref{eq:DbracketonG} with the canonical Dorfman bracket in generalized geometry \cite{Hitchin:2004ut,Gualtieri:2003dx}, it is instructive to study maps\footnote{In \cite{Freidel:2017yuv} this map is denoted by $\rho: L\oplus L^*\rightarrow T\PS$. Here we use $\widehat{E}$ because on a group manifold this map is closely related to the twist matrix of generalized Scherk-Schwarz reductions \cite{Grana:2012rr,Geissbuhler:2011mx,Aldazabal:2011nj}.} $\widehat{E}$ from $T \mathcal{M} \oplus T \mathcal{M}^*$ to $T \PS$. Following \cite{Freidel:2017yuv}, $\mathcal{M}$ denotes the partition $\mathcal{M} = \coprod_{[\ell]} M_\ell$ of $\PS$ as a set of leafs $M_\ell \in \mathcal{L}\backslash\PS$ where the index space is the Lie group $\mathcal{L}$. If we view the foliation $\mathcal{M}$ as a $d$-dimensional manifold, we can define its tangent and co-tangent bundle. The restriction of $T \mathcal{M} \oplus T^* \mathcal{M}$ to any leaf $M_\ell$ is equivalent to the generalized tangent bundle of the target space $M$.

On $T \mathcal{M} \oplus T^* \mathcal{M}$ we introduce the canonical pairing
\begin{equation}\label{eqn:hateta}
  \widehat{\eta}( x + \phi, y + \xi ) = \phi(y) + \xi(x)\,, \quad x, y \in T \mathcal{M}\,,\,\, \phi, \xi \in T^* \mathcal{M} \,.
\end{equation}
We define $\widehat{E}$ such that the two relations
\begin{equation}
  \widehat{E} x = x \quad \text{and} \quad
  \eta\left( \widehat{E} (x + \phi), \widehat{E} (y + \xi) \right) = \widehat{\eta}( x + \phi, y + \xi)
\end{equation}
hold. However, they do not fix $\widehat{E}$ completely. To do so, we further impose
\begin{equation}\label{eqn:hatomega}
  \omega \left( \widehat{E} (x + \phi), \widehat{E} (y + \xi) \right) = \widehat{\omega}( x + \phi, y + \xi)
  \quad \text{with} \quad \widehat{\omega}( x + \phi, y + \xi ) = \phi(y) - \xi(x)\,.
\end{equation}
Under this map, the D-bracket $\bl \, , \br^\nabla$ reduces on each leaf of $\mathcal{M}$ to the canonical Dorfman-bracket
\begin{equation}\label{eqn:canonicalDB}
  \bl \widehat{E}(x + \phi), \widehat{E}(y + \xi) \br^\nabla \widehat{E}^{-1} =
  \bl x+\phi, y+\xi \br = [x, y] + L_x \xi - \iota_y \dd \phi \,
\end{equation}
where $L_x$ is the ordinary Lie derivative on $M$. This proves the claim that \eqref{eq:DbracketonG} is a unique D-bracket. The full calculation is straightforward but cumbersome. If both $L$ and $\Lt$ are integrable, it follows directly the proof of proposition 1 in \cite{Freidel:2017yuv}. Instead of presenting the details here, we rather note that this relation is equivalent to
\begin{equation}\label{eqn:framealgebra}
  \begin{aligned}
    \widehat{\eta}\Big( \bl \iota_A \widehat{E}^{-1},  \iota_B \widehat{E}^{-1} \br^\nabla , \iota_C \widehat{E}^{-1}(Z) \Big)  &= \eta( \bl E^{-1} T_A , E^{-1} T_B \br, E^{-1} T_C ) =  \iota_C \iota_B \iota_A F \\
      & = \widehat{\eta} \Big( \iota_A \dd \iota_B \widehat{E}^{-1} , \iota_C \widehat{E}^{-1} \Big) + \text{cycl.}
  \end{aligned}
\end{equation}
with $\iota_A$ as an abbreviation for $\iota_{E^{-1} T_A}$. This relation is essential for generalized Scherk-Schwarz \cite{Grana:2012rr,Geissbuhler:2011mx,Aldazabal:2011nj} reductions in DFT. There  $\iota_A \widehat{E}^{-1}$ is called twist matrix or generalized frame field.  Before we further explore this connection in the next subsection, we first prove the frame algebra \eqref{eqn:framealgebra}:
\begin{proof}
We start with the identity
\begin{equation}\label{eqn:domegaproof}
  \dd \omega = \frac12 \, \dd \widehat{\omega}( \widehat{E}^{-1} \cwedge \widehat{E}^{-1} ) =
  \widehat{\omega} ( \dd \widehat{E}^{-1} \cwedge \widehat{E}^{-1} ) = F \,.
\end{equation}
We are allowed to pull $\dd$ into $\widehat{\omega}$ because according to \eqref{eqn:hatomega} it does not have any intrinsic coordinate dependence. Furthermore, $\widehat{E}^{-1}$ is just the identity on $T \mathcal{M}$ and therefore the image of $\dd \widehat{E}^{-1}$ is restricted to $T^* \mathcal{M}$. Thus, $ ( \widehat{\eta} + \widehat{\omega} ) ( d \widehat{E}^{-1} \cwedge \widehat{E}^{-1} ) = 0$ and we can also write
\begin{equation}
  - \widehat{\eta}( d \widehat{E}^{-1} \cwedge \widehat{E}^{-1} ) = F \,.
\end{equation}
Finally, we apply $\iota_C \iota_B \iota_A$ to both sides of this equation, resulting in
\begin{equation}
  \iota_{X_A} \iota_{X_B} \iota_{X_C} \widehat{\eta} ( \dd \widehat{E}^{-1} \cwedge \widehat{E}^{-1} ) =
  \widehat{\eta} ( \iota_{X_A} \iota_{X_B} \dd \widehat{E}^{-1} , \iota_{X_C} \widehat{E}^{-1} ) + \text{cycl.} = \iota_C \iota_B \iota_A F \,.
\end{equation}
Now we are almost there, the only thing we have to do is to swap the exterior derivative $\dd$ and the interior product $\iota_B$ in front of $\widehat{E}^{-1}$. For this purpose the Lie derivative
\begin{equation}
  L_{X_A} \widehat{E}^{-1} = \iota_{X_A} \dd ( \iota_{X_B} \widehat{E}^{-1} ) \wedge E^B - F_{AB}{}^C E^B \iota_{X_C} \widehat{E}^{-1}
\end{equation}
is helpful ($E^A$ is $T_A E^A = p^{-1} \dd p$). Using this identity, we finally find
\begin{equation}
  \widehat{\eta}( \iota_A d ( \iota_B \widehat{E}^{-1} ) , \iota_C \widehat{E}^{-1} ) + \text{cycl.} =
    \iota_C \iota_B \iota_A F \,.
\end{equation}
This proves \eqref{eqn:framealgebra}.
\end{proof}

Physically, the D-structure captures the global symmetries of the worldsheet theory. A way to prove this claim is to apply $\widehat{E}$ to the $\sigma$-model in \eqref{eqn:Ssigma}. In particular, we define
\begin{equation}
  x_\sigma + \phi_\sigma = \widehat{E}^{-1} X_\sigma \quad \text{and} \quad
  x_\tau + \phi_\tau = \widehat{E}^{-1} X_\tau
\end{equation}
and obtain
\begin{equation}\label{eqn:Ssigma2}
  S = \int \dd\sigma \,\dd\tau \left( \mathtt{p} ( x_\tau ) - \frac12 \widehat{\mathcal{H}}( x_\sigma + \mathtt{p}, x_\sigma + \mathtt{p} )\right)
\end{equation}
after the identification $\mathtt{p} = \phi_\sigma$. The most interesting part of this equation is  the generalized metric $\widehat{\mathcal{H}}$. It is defined in the same way as $\widehat{\eta}$ and $\widehat{\omega}$,
\begin{equation}\label{eqn:genframealgebra}
  \mathcal{H}\left( \widehat{E}(x + \phi), \widehat{E}(y + \xi) \right) = \widehat{\mathcal{H}}(x + \phi, y + \xi)\,.
\end{equation}
A convenient parameterization for this symmetric, rank-two tensor is
\begin{equation}\label{eqn:genmetricGB}
  \widehat{\mathcal{H}}(x + \phi, y + \xi) = \begin{pmatrix} x^i & \phi_i \end{pmatrix}
    \begin{pmatrix}
      G_{ij} - B_{ik} G^{kl} B_{lj} & B_{ik} G^{kj} \\
      - G^{ik} B_{kj} & G^{ij}
    \end{pmatrix}
  \begin{pmatrix} y^j \\ \xi_j \end{pmatrix}\,.
\end{equation}
Plugging it into the action \eqref{eqn:Ssigma2} and furthermore assuming the it just depends on the physical coordinates $m$ but not on $\ell$, we are able to integrate out $\mathtt{p}$ to find
\begin{equation}\label{eqn:piofGB}
  \mathtt{p}_i = G_{ij} x^i_\tau + B_{ij} x^j_\sigma\,. 
\end{equation}
This gives rise to the canonical $\sigma$-model action
\begin{equation}
  S = \frac12 \int \dd\sigma\, \dd\tau \Big( G( x_\tau, x_\tau ) - G( x_\sigma, x_\sigma ) + 2 B( x_\tau, x_\sigma ) \Big)\,.
\end{equation}
Its global symmetries are encompassed by diffeomorphisms and $B$-field gauge transformations on the target space. Both can be combined into generalized diffeomorphisms whose infinitesimal version is mediated by the generalized Lie derivative which is related to the D-bracket via \eqref{eqn:genLie}.  Applying it to the generalized metric gives rise to
\begin{equation}
  \delta \widehat{\mathcal{H}} = \genlie_{x+\phi} \widehat{\mathcal{H}}
    \qquad \leftrightarrow \qquad
    \delta G = L_x G \quad \text{and} \quad \delta B = L_x B + d \phi
\end{equation}
if we use the parameterization \eqref{eqn:genmetricGB} and the canonical $D$-bracket in \eqref{eqn:canonicalDB}. From this discussion, we see that the Poisson-Lie $\sigma$-model and the corresponding para-Hermitian structure are connected in an intriguing way.

\subsubsection*{Double Field Theory on group manifolds}
The low energy effective target space theory of the $\mathcal{E}$-model \eqref{eqn:Ssigma2} is supergravity. However, in supergravity some of its features, like Poisson-Lie T-duality, are obscured. An equivalent description which makes them manifest is Double Field Theory on group manifolds \cite{Blumenhagen:2014gva,Blumenhagen:2015zma,Hassler:2017yza}, or \DFTwzw{} for short\footnote{For a recent review see \cite{Demulder:2019bha}.}. Since the almost para-Hermitian structure is essential for the $\sigma$-model, it should also appear there. The map in table~\ref{tab:DFTwzwobjects} presents the explicit relations.
\begin{table}[!t]
\begin{center}
\renewcommand{\arraystretch}{1.5}
\begin{tabular}[]{lc}
\toprule
  $\eta$-metric & $\eta_{AB} = \langle T_A, T_B \rangle$ \\
  generalized metric & $\mathcal{H}_{AB} = \langle T_A, \mathcal{E} T_B \rangle$ \\
  structure coefficients & $F_{ABC} = \langle [ T_A, T_B ], T_C \rangle$ \\
  flat derivative & $D_A = \iota_A D$  \\
  covariant derivative & $\nabla_A = \iota_A \nabla$ \\
  generalized Lie derivative & $\genlie_{\xi} V_A =  \iota_A \bl X_B \xi^B, X_C V^C \br^\nabla$ \\
  section condition & $N_K(X,Y) = 0$ \\
  generalized frame field & $\widehat{E}_A{}^{\hat I} \Big( \, \phi_i \,\,\, x^i \, \Big) = 
    ( x + \phi ) \iota_A \widehat{E}^{-1}$ \\ \bottomrule
\end{tabular} 
\end{center}
\caption{Relation between the crucial ingredients for the half-integrable structure on group manifolds and \DFTwzw{}. To simplify the notation for the generalized Lie derivative, we use the vector fields $X_A = E^{-1} T_A$ which generate right translations on $\PS$.}\label{tab:DFTwzwobjects}
\end{table}
Calculating $\iota_A \widehat{E}$ explicitly, we obtain
\begin{equation}\label{eqn:Ehatinv}
  \iota_A \widehat{E}^{-1} = x_A + \underbrace{\langle m^{-1} \dd m, T_A \rangle - \frac12 \langle m^{-1} \dd m, \iota_{x_A} m^{-1} \dd m \rangle}_{\displaystyle\varphi_A} - \frac12 \iota_{x_A} \BW \,,
\end{equation}
where the contribution $\varphi_A$ is significant for the discussion in section~\ref{sec:dressingGG}. Here $x_A$ denotes the push forward $x_A = \pi^* E^{-1} T_A$ where $\pi$ projects from $\PS$ to $M$ ($\pi: \PS \rightarrow M$). If we additionally impose the constraint \eqref{eqn:addrestriction} the second term of $\varphi_A$ vanishes and the frame field is equivalent to the one discussed by \cite{Demulder:2018lmj} in the context of \DFTwzw{}.

Let us finally rewrite the frame algebra \eqref{eqn:framealgebra} that we proved above in the language of DFT. Applying the dictionary in table~\ref{tab:DFTwzwobjects} we obtain
\begin{equation}
  3 \widehat{E}_{[A}{}^{\hat I} \partial_{\hat I} \widehat{E}_B{}^{\hat J} \widehat{E}_{C]\hat J} = F_{ABC}\,,
\end{equation}
where the generalized frame field has the two additional properties:
\begin{itemize}
  \item It just depends to the physical directions $m$.
  \item It transforms $\eta_{AB}$ to the canonical form
    \begin{equation}
      \widehat{E}^A{}_{\hat I} \eta_{AB} \widehat{E}^B{}_{\hat J} = \widehat{\eta}_{\hat I\hat J} =
        \begin{pmatrix}
          0 & \delta^i_j \\
          \delta_i^j & 0
        \end{pmatrix}
    \end{equation}
    which is also used to lower hatted indices.
\end{itemize}
Hence it is equivalent to the twist matrix used for generalized Scherk-Schwarz reductions \cite{Grana:2012rr,Geissbuhler:2011mx,Aldazabal:2011nj}. They automatically give rise to consistent truncations.

\subsection{Born geometry}
The generalized frame field $\widehat{E}$ is also the perfect tool to eventually make the connection with Born geometry introduced in section~\ref{sec:BornGeometry}. A Born structure on $\PS$ relies on the algebraic constraints in table~\ref{tab:BornGeo} which have to be imposed on $\eta$, $\omega$ and $\mathcal{H}$. However there are no derivatives involved. Thus, these constraints are invariant under arbitrary, in general coordinate depended, GL($2d$) transformations. To solve them, it is convenient to find a transformation which brings $\eta$ and $\omega$ into the canonical form in equation \eqref{m}. But as we see from \eqref{eqn:hateta} and \eqref{eqn:hatomega}, the generalized frame field $\widehat{E}$ takes by construction care of this task. Thus, we conclude that the generalized metric $\widehat{\mathcal{H}}$, defined in \eqref{eqn:genmetricGB}, should have a vanishing $B$-field in order to give rise to a Born geometry. If we have a background with a $B$-field that is pure gauge and therefore satisfies $\dd B = 0$, we can adsorb it into $\BW$ in the definition of $\omega$ \eqref{eqn:omegaXY}. Hence we conclude that Born geometries on $\PS$ just arise if there is no $H$-flux on the target space.

\subsection{Poisson-Lie symmetry, integrability and T-duality}\label{sec:isometries}
There are four tensors on the group manifold $\PS$ which are essential to our discussion in this section, $\eta$, $\omega$, $\HH$ and $F$. We have seen that they are the building blocks of Poisson-Lie $\sigma$-models. Now we want to have a closer look at their isometries. Isometries are generated by Killing vectors. The maximal isometry group on $\PS$ is $\PS_{\mathrm{L}} \times \PS_{\mathrm{R}}$. The first factor captures translations by left action of a group element and the second one by right action. Infinitesimally those translations are generated by the killing vectors
\begin{equation}
  \xi^{\mathrm{L}}_A = V^{-1}(T_A) \quad \text{and} \quad \xi^{\mathrm{R}}_A = E^{-1}(T_A)\,.
 \label{eq:KillingVectors}
\end{equation}
$E^{-1}$ and $T_A$ we already know, while $V^{-1}$ is the inverse of the right-invariant Maurer-Cartan form
\begin{equation}
  V(X) = \dd p p^{-1} (X)\,.
\end{equation}
The Lie algebra corresponding to the maximal isometry group is generated by
\begin{align}
  L_{\xi^{\mathrm{L}}_A} E(X) & = 0 & 
  L_{\xi^{\mathrm{R}}_A} E(X) & = - [ T_A, E(X) ] \\
  L_{\xi^{\mathrm{L}}_A} V(X) & = [ T_A, V(X) ] &
  L_{\xi^{\mathrm{R}}_A} V(X) & = 0 \,.
\end{align}
Using these relations it is not hard to show that
\begin{equation}
  L_{\xi^{\mathrm{L}}_A} \eta = L_{\xi^{\mathrm{R}}_A} \eta = 0 \quad \text{and} \quad
  L_{\xi^{\mathrm{L}}_A} F = L_{\xi^{\mathrm{R}}_A} F = 0 
\end{equation}
hold. Thus, both $\eta$ and $F$ are bi-invariant. The generalized metric can in general break all the isometries. But most interesting are cases where it preserves some of them. If they are freely acting (without any fixed point), and the isometry group has all the properties we discussed for $\PS$, $\mathcal{H}$ admits Poisson-Lie symmetry. In the following, we assume without loss of generality this isometry group to be $\PS_{\mathrm{L}}$.

Poisson-Lie symmetry is governed by a ``non-commutative conservation law'' \cite{Klimcik:1995ux}. To see how this is connected to the isometry group $\PS_{\mathrm{L}}$ of the generalized metric, we vary the $\sigma$-model action \eqref{eqn:Ssigma} with respect to a small change of the coordinates $\delta X = E^{-1} T_A \delta \epsilon^A(\tau, \sigma)$ and require the variation to vanish,
\begin{equation}\label{eqn:deltaS}
  \delta S = - \frac12 \int \dd\sigma \,\dd\tau \delta \epsilon^A \genlie_{\xi_A^{\mathrm{L}}}^\nabla \mathcal{H}(X_\sigma, X_\sigma) - \int \delta \epsilon^A \dd J_A = 0 \,.
\end{equation}
This is the standard Noether procedure to identify conserved currents. Since left invariant vector fields are covariant constant under the flat derivative $D$ ($D_X \xi^{\mathrm{L}}_A = 0$), a generalized metric with $\PS$ as left isometry group transforms as
\begin{equation}
  \genlie_{\xi^{\mathrm{L}}_A}^\nabla \mathcal{H}(X_\sigma, X_\sigma) = 2 F( \xi^{\mathrm{L}}_A, \mathcal{E} X_\sigma, X_\sigma)
\end{equation}
under the generalized Lie derivative. This allows us to constraint the current $J_A$ by imposing that the variation of the action in \eqref{eqn:deltaS} vanishes:
\begin{equation}\label{eqn:flatCurrent}
  \dd J = - [\mathcal{E} X_\sigma, X_\sigma ] \, \dd \tau \wedge \dd \sigma\,.
\end{equation}
In this equation we use $\mathcal{E}$ instead of the generalized metric $\mathcal{H}$. They are related by \eqref{eqn:EtoH}. To obtain $J$, we have to integrate this relation. In general this is not possible unless $J$ is a flat $\PS$-connection on the worldsheet. Flatness requires it to satisfy the Maurer-Cartan equation
\begin{equation}\label{eqn:flatCurrent2}
  \dd J + \frac12 J \wedge J = 0\,,
\end{equation}
in this context also called non-commutative conservation law \cite{Klimcik:1995ux}. One can bring the current that solves this equation (and also arises from the variation of the action) in a very suggestive form. The worldsheet coordinates $\tau$ and $\sigma$ are obstructing its structure a bit. Hence, it is better to use light-cone coordinates instead,
\begin{equation}
  \xi^\pm = \frac12(\tau \pm \sigma) \,, \quad \partial_\tau = \frac12 (\partial_+ + \partial_-) \quad \text{and} \quad
  \partial_\sigma = \frac12 (\partial_+ - \partial_-)\,.
\end{equation}
With this notation the current reads
\begin{equation}
  J = ( \mathcal{E} + 1 ) X_\sigma \dd \xi^+ + ( \mathcal{E} - 1 ) X_\sigma \dd \xi^- = J_+ \dd \xi^+ + J_- \dd \xi^-
\end{equation}
and one can easily check that it satisfies the equations \eqref{eqn:flatCurrent} and \eqref{eqn:flatCurrent2}\,. Restricting to just the current components which capture small changes of the physical coordinates, we find
\begin{equation}
  J \widehat{E}( v_a ) = v_a{}^i (G_{ij} + B_{ij}) \partial_+ x^j \dd \xi^+  - v_a{}^i (G_{ij} - B_{ij}) \partial_- x^j \dd \xi^- 
\end{equation}
after expressing $\mathcal{E}$ in terms of the generalized metric \eqref{eqn:genmetricGB} and substitution \eqref{eqn:piofGB}. This expression is equivalent to the current presented in \cite{Klimcik:1995ux} assuming that the target space $M$ is a group manifold and that $v_a$ denotes its left-invariant vector fields. 

For a particular class of $\mathcal{E}$-models, the current $J$ admits a decomposition into two contributions $\mathcal{R}$ and $\mathcal{J}$ valued in the $d$-dimensional real Lie algebra $\mathfrak{g}$. For them the conservation law \eqref{eqn:flatCurrent} takes the simple form
\begin{equation}\label{eqn:eomPCM}
  \begin{aligned}
    \partial_\tau \mathcal{R} &= \partial_\sigma \mathcal{J} + [ \mathcal{J}, \mathcal{R} ]_{\mathfrak{g}} \\
    \partial_\tau \mathcal{J} &= \partial_\sigma \mathcal{R}
  \end{aligned}
\end{equation}
where $\brac_{\mathfrak{g}}$ denotes the Lie bracket of $\mathfrak{g}$. These are the Zakharov-Mikhailov field equations for the principal chiral model \cite{zakharov1978relativistically}. They can be rewritten in terms of a family of flat connections labeled by the spectral parameter $\lambda \in \mathcal{C}/\{\pm 1\}$
\begin{equation}
  \mathbb{A}_\pm( \lambda ) = \frac{\mathcal{J} \pm \mathcal{R}}{1 \pm \lambda}
  \quad \text{satisfying} \quad
  \partial_+ \mathbb{A}_- (\lambda) - \partial_- \mathbb{A}_+ (\lambda) + [ \mathbb{A}_-(\lambda), \mathbb{A}_+(\lambda) ] = 0\,.
\end{equation}
It allows us to derive an infinite number of conserved charges and thus showing that the $\sigma$-model is integrable.

\section{Dressing cosets}\label{sec:dressingcoset}
In the first part of this paper we connected a mathematical structure to its physical applications. Now we follow the opposite approach and instead use the dressing coset construction for Poisson-Lie $\sigma$-models to explore a vast class of new examples of para-Hermitian geometries. 

\subsection{Gauged Poisson-Lie \texorpdfstring{$\sigma$}{sigma}-model}
From a physical point of view, the isometries discussed in section~\ref{sec:isometries} represent global symmetries of the worldsheet $\sigma$-model. If there are no obstructions, one can promote them to local symmetries by gauging. Since the action~\eqref{eqn:Ssigma} contains a WZ-term, the gauged action does not just involve minimal coupling, but one has to be a bit more careful. In general not every subgroup of $\PS_{\mathrm{L}} \times \PS_{\mathrm{R}}$ is suitable to be gauged. Hull and Spence show in \cite{Hull:1989jk,Hull:1990ms} (see also \cite{Jack:1989ne}) the constraints it has to satisfy to permit gauging. 

Before we discuss these conditions, let us fix the notation for this section. The isometries we want to gauge are specified by a set of Killing vectors $\xi_\alpha$ where the index $\alpha$ (a subset of the index $A$) labels the isometries. They form a subgroup $\mathcal{F}$ with generators $T_\alpha$ and structure constants $f_{\alpha\beta}{}^\gamma$. Now the constraints from \cite{Hull:1990ms} can be stated as follows:

\begin{itemize}
  \item First, there has to be a set of globally defined one-forms $A_\alpha$ fulfilling    \begin{equation}\label{eqn:dva}
      \iota_\alpha F = - \dd A_\alpha \,,
    \end{equation}
    where we use the abbreviation $\iota_\alpha = \iota_{\xi_\alpha}$. This constraint is equivalent to $F$, the closed three-form in \eqref{eq:F} or the structure coefficients of $\PS$, being invariant under the action of $\xi_\alpha$.
  \item Second, the $A_\alpha$ obtained in this way has to give rise to the Lie algebra of the gauge group
    \begin{equation}\label{eqn:Lavb}
      L_\alpha A_\beta = f_{\alpha\beta}{}^\gamma A_\gamma \,.
    \end{equation}
  \item Finally, the combination $\iota_\alpha A_\beta$ has to be skew-symmetric.
\end{itemize}
We want to keep the Poisson-Lie symmetry of the $\sigma$-model intact. Thus, we do not touch the left isometry group and only consider subgroups of $\PS_{\mathrm{R}}$ for gauging. In this case, \eqref{eqn:dva} gives rise to
\begin{equation}
  \iota_\alpha F = - \dd A_\alpha = - \dd \langle T_\alpha , p^{-1} \dd p \rangle \,.
\end{equation}
While the second constraint \eqref{eqn:Lavb} is automatically fulfilled, the third one requires
\begin{equation}
  \iota_\alpha A_\beta = \langle T_\alpha, T_\beta \rangle = 0
\end{equation}
and therefore tells us that the subgroup $\mathcal{F}$ of $\PS_{\mathrm{R}}$ we gauge has to be isotropic. Otherwise $\iota_\alpha A_\beta$ would have a non-trivial symmetric contribution. Intriguingly this is exactly the same restriction which arises in the dressing coset construction presented in \cite{Klimcik:1996np}.

The idea for the remainder of this section is to use the physical guidance from the gauged WZW-model to obtain a large family of examples for the mathematical structures presented above. To gauge the action, we introduce the worldsheet connection $\hat A^\alpha_\mu$ where $x^\mu=(\tau,\sigma)$ are the worldsheet coordinates. In two dimensions gauge fields are not dynamical. Therefore we can integrate them out. For WZW-models based on a compact Lie group with a negative definite pairing $\langle \ , \ \rangle$, the action is quadratic in $\hat A^\alpha_\mu$ and integrating it out results in a particular gauge fixing. But for the $\mathcal{E}$-model the pairing is indefinite and in connection with $\mathcal{F}$ being restricted to an isotropic subgroup of $\PS$, the gauged action is just linear in $\hat A^\alpha_\mu$. Thus the worldsheet connection plays the role of a Lagrange multiplier and restricts the dynamical fields by
\begin{equation}
  \eta( \xi^{\mathrm{R}}_\alpha, X_\mu) = \langle T_\alpha , X_\mu \rangle = 0\,,
\end{equation}
where $X_\mu=(X_\tau,X_\sigma)$ are the two vectors in \eqref{eqn:XsigmaXtau} which specify the embedding of the worldsheet in $T\PS$ and $\xi^{\mathrm{R}}_\alpha$ are the Killing vectors generating $\PS_{\mathrm{R}}$. They are defined in \eqref{eq:KillingVectors}. The constraint states that only embeddings orthogonal to the isometry vectors $\xi^{\mathrm{R}}_\alpha$ are to be considered. Not surprisingly, this is the same constraint which was imposed in \cite{Klimcik:1996np} to obtain the dressing cosets.

\subsection{Horizontal almost para-Hermitian structure}
From the target space perspective, this constraint can be conveniently implemented by restricting to the horizontal subspace defined by the connection one-form
\begin{equation}
  A^\alpha(X) =  \eta( E T^\alpha,  X )\,.
  \label{eq:TSconnection}
\end{equation}
Note that this is a connection on $\PS$, with $A^\alpha=A^\alpha_AE^A$, as opposed to the connection on the worldsheet $\hat A^\alpha_\mu$. Here $T^\alpha$ is chosen such that it is dual to $T_\alpha$ with respect to $\eta$ which implies
\begin{equation}
  \langle T^\alpha, T^\beta \rangle = 0 \quad \text{and} \quad
  \langle T^\alpha, T_\beta \rangle = \delta^\alpha_\beta\,.
\end{equation}
Thus the generators $T^\alpha$ are isotropic as well. Still they are not on the same footing as $T_\alpha$ because they do not close into a Lie algebra. For later convenience we will combine them into $T^{\underline{A}}=\begin{pmatrix} T_\alpha & T^\alpha \end{pmatrix}$. Now we ready to define the projectors onto the vertical and horizontal subspaces
\begin{equation}
  \PdcV (X) := \xi^{\mathrm{R}}_{\alpha} A^{\alpha}(X) + \xi^{\mathrm{R}\alpha} A_{\alpha}(X) = \xi^{\mathrm{R}\underline{A}}A_{\underline{A}}(X) \quad \text{and} \quad
  \PdcH(X) := X - \PdcV(X) \, .
  \label{eq:DCprojectors}
\end{equation}
From the definition \eqref{eq:DCprojectors} one can show that $\eta(\PdcV(X),Y) = \eta(X,\PdcV(Y))$ which further implies $\eta(\PdcH(X),Y) = \eta(X,\PdcH(Y))$ or simply $\PdcH^\T\eta=\eta\PdcH$.

The idea is now to restrict all quantities which we have discussed in the context of para-Hermitian and Born geometry to the horizontal subspace by applying $\PdcH$. We do not really care about the vertical contributions because they just capture the gauge symmetry of the model. For $\eta$ and $\omega$ this idea gives rise to
\begin{align}
  \eta_{\mathtt{h}} (X, Y) &= \eta( \PdcH X, \PdcH Y )\,, \\
  \omega_{\mathtt{h}} (X, Y) &= \omega(  \PdcH X, \PdcH Y )\,.
\end{align}
On the horizontal subspace $\eta_{\mathtt{h}}$ is invertible, its inverse is given by $ \eta^{-1}_{\mathtt{h}} = \PdcH \eta^{-1} \PdcH^\T $ and satisfies $  \eta^{-1}_{\mathtt{h}} \eta_{\mathtt{h}} = \PdcH$ where $\PdcH$ acts as the ``identity'' on the horizontal subspace. The same statement holds for $\omega_{\mathtt{h}}$. Thus, we are able to construct
\begin{equation}\label{eqn:Khorizontal}
  K_{\mathtt{h}} = \eta^{-1}_{\mathtt{h}} \omega_{\mathtt{h}} \quad \text{with} \quad K_{\mathtt{h}}^2 = \PdcH
\end{equation}
and the projectors
\begin{align}
  P_{\mathtt{h}} &= \frac12 ( \PdcH + K_{\mathtt{h}} ) = \PdcH P\PdcH\,, \\
  \Pt_{\mathtt{h}} &= \frac12 ( \PdcH - K_{\mathtt{h}} ) = \PdcH  \Pt \PdcH\,.
\end{align}
They add up to $\PdcH$ and implement an almost para-Hermitian structure on the horizontal subspace.

\subsection*{Generalized geometry on $T M \oplus T^* M$}\label{sec:dressingGG}
As for the plain group manifold case in the last section, we want to make contact with the generalized geometry of the $\sigma$-model's target space. To this end, we again use the map $\widehat{E}$ and its inverse $\widehat{E}^{-1}$. Like for $\eta$ and $\omega$, we are only interested in the restriction to the horizontal subspace
\begin{equation}
  \widehat{E}^{-1}_{\mathtt{h}} = \widehat{E}^{-1} \PdcH \qquad \text{and} \qquad
  \widehat{E}_{\mathtt{h}} = \PdcH \widehat{E} \,.
\end{equation}
Furthermore, we have to introduce the projected versions of $\widehat{\eta}$ and $\widehat{\omega}$ which we defined in \eqref{eqn:hateta} and \eqref{eqn:hatomega}. The natural generalization of these two quantities is
\begin{equation}\label{eqn:etahandomegah}
  \begin{aligned}
    \widehat{\eta}_{\mathtt{h}}(\widehat{E}^{-1}_{\mathtt{h}} X, \widehat{E}^{-1}_{\mathtt{h}} Y) &= \eta_{\mathtt{h}}(X, Y)\,, \\
  \widehat{\omega}_{\mathtt{h}}(\widehat{E}^{-1}_{\mathtt{h}} X, \widehat{E}^{-1}_{\mathtt{h}} Y) &= \omega_{\mathtt{h}}(X,Y)\,.
  \end{aligned}
\end{equation}
In contrast to the non-degenerate case in section~\ref{sec:Dstructure2}, these relations do not fix $\widehat{\eta}_{\mathtt{h}}$ and $\widehat{\omega}_{\mathtt{h}}$ completely. Thus, we impose additionally that both vanish in the directions corresponding to the gauge symmetry transformations of the gauged $\sigma$-model
\begin{equation}
  \widehat{\eta}_{\mathtt{h}}(x_\alpha, y + \xi) = 0 \qquad \text{and} \qquad
  \widehat{\omega}_{\mathtt{h}}(x_\alpha, y + \xi) = 0 \qquad \forall\, y \in T \mathcal{M}\,, \quad
    \xi \in T^* \mathcal{M}\,.
\end{equation}
The Killing vector $x_\alpha$ generates these transformations which we discussed above in details. It is defined on each leaf of the foliation $\mathcal{M}$ by the pullback of $\xi_\alpha$ and is also an important part of the generalized frame field $\widehat{E}^{-1}$ in \eqref{eqn:Ehatinv}. However this particular choice is in general not compatible with \eqref{eqn:etahandomegah}. It requires additionally
\begin{equation}\label{eqn:ialphaEinv}
  \iota_\alpha \widehat{E}^{-1} = x_\alpha
\end{equation}
to hold. According to \eqref{eqn:Ehatinv} this constraint is equivalent to
\begin{equation}\label{eqn:restrBwzw}
  \iota_\alpha \BW = 2 \varphi_\alpha\,.
\end{equation}
We will discuss the intriguing consequences it implies later. For the moment we rather establish a frame algebra analogous to \eqref{eqn:framealgebra}, namely
\begin{equation}\label{eqn:framealgebraDC}
  \widehat{\eta}_{\mathtt{h}}\left( \iota_{\bar A} \dd \iota_{\bar B} \widehat{E}^{-1}_{\mathtt{h}} , \iota_{\bar C}   \widehat{E}^{-1}_{\mathtt{h}} \right) + \text{cycl.} = \iota_{\bar C} \iota_{\bar B} \iota_{\bar A} F\,.
\end{equation}
The overbarred indices $\bar A, \bar B, \dots$ are the complement to the underbarred indices $\underline{A}, \underline{B},\dots$ and exclude the generators $T^{\underline{A}}$ which generate translations along the vertical subspace. We therefore have $T^A = (T^{\underline{A}}, T^{\bar A})$.
\begin{proof}
We start with the analog to \eqref{eqn:domegaproof}
\begin{equation}
	\left. \dd \omega_{\mathtt{h}} \right|_{\mathtt{h}} = F_{\mathtt{h}} + 2 \varphi^\alpha \wedge \dd A_\alpha = \left. \widehat{\omega}_{\mathtt{h}}( \dd \widehat{E}^{-1}_{\mathtt{h}} , \widehat{E}^{-1}_{\mathtt{h}} ) \right|_{\mathtt{h}}\,.
\end{equation}
Here $|_{\mathtt{h}}$ denotes the restriction of a differential form to the horizontal subspace and $F_{\mathtt{h}}(X,Y,Z)=F(\PdcH X, \PdcH Y,\PdcH Z)$. Next we evaluate
\begin{equation}
	\left. ( \widehat{\eta}_{\mathtt{h}} + \widehat{\omega}_{\mathtt{h}} ) ( \dd \widehat{E}^{-1}_{\mathtt{h}} ,  \widehat{E}^{-1}_{\mathtt{h}} )\right|_{\mathtt{h}} = - 2 \varphi^\alpha \wedge \dd A_\alpha\,.
\end{equation}
All remaining steps are the same as for the group manifolds case discussed in section~\ref{sec:Dstructure}.
\end{proof}

Using \eqref{eqn:framealgebraDC}, we can show that the canonical D-bracket arises after applying $\widehat{E}_{\mathtt{h}}$ and $\widehat{E}^{-1}_{\mathtt{h}}$to the D-structure on $\PS$,
\begin{equation}\label{eqn:projectedgenLie}
  \bl \widehat{E}_{\mathtt{h}}(x + \phi), \widehat{E}_{\mathtt{h}}(y + \xi) \br^\nabla \widehat{E}^{-1}_{\mathtt{h}} = \bl \widehat{\Pi}_{\mathtt{h}} (x+\phi), \widehat{\Pi}_{\mathtt{h}}( y+\xi )\br \widehat{\Pi}_{\mathtt{h}}^\T \,.
\end{equation}
$\bl \, , \, \br$ denotes the Dorfman bracket defined in \eqref{eqn:canonicalDB} and $\widehat{\Pi}_{\mathtt{h}}$ is the projector to the horizontal subspace on the generalized tangent bundle
\begin{equation}
  \widehat{\Pi}_{\mathtt{h}} = \widehat{E}^{-1} \PdcH \widehat{E}\,.
\end{equation}
This bracket does not close in general and we have to additionally impose
\begin{equation}
  \bl \widehat{\Pi}_{\mathtt{h}} (x+\phi), \widehat{\Pi}_{\mathtt{h}}( y+\xi )\br \widehat{\Pi}_{\mathtt{v}}^\T = 0
    \quad \text{with} \quad
  \widehat{\Pi}_{\mathtt{v}} = 1 - \widehat{\Pi}_{\mathtt{h}} \,.
\end{equation}
for its closure. Due to \eqref{eqn:ialphaEinv}, this condition is equivalent to
\begin{equation}
  \bl x_\alpha , \widehat{\Pi}_{\mathtt{h}}( y+\xi )\br \widehat{\Pi}_{\mathtt{h}}^\T = 0
\end{equation}
which tells us that the generalized vector $y+\xi$ has to be invariant under the symmetry generated by $x_\alpha$. But this was exactly our starting point. In the dressing coset construction, we imposed from the beginning that this transformations leaves all quantities in the $\sigma$-model invariant. Otherwise it would not have been possible to carry out the gauging. Respecting this condition, we eventually can restrict the discussion to the dressing coset $N=\mathcal{L}\backslash\PS/\mathcal{F}$. An explicit parametrization of the coset representative $m$ is
\begin{equation}
	M \ni m = n f \qquad f \in \mathcal{F} \qquad
	\text{and} \qquad n \in N\,,
\end{equation}
where $\mathcal{F}$ denotes the subgroup we gauge. It gives rise to two maps $\pi$ and $\sigma$:
\begin{equation}
\tikz[baseline=-2pt]{
  \matrix (m) [matrix of math nodes, column sep=3em] {%
     M &  N  \\ };
  \draw[->,shiftup] (m-1-1.east) -- (m-1-2.west |- m-1-1) node[midway, above] {$\pi$};
  \draw[<-,shiftdn] (m-1-1.east) -- (m-1-2.west |- m-1-1) node[midway, below] {$\sigma$};}
\quad \text{with} \quad
\pi ( n, f ) = n \quad \text{and} \quad
\sigma (n)  = n f_0, \quad f_0 \in  \mathcal{F} \, .
\end{equation}
Here $f_0$ is an arbitrary but constant element in $ \mathcal{F}$, meaning it should not change with $n$. A canonical choice is the identity $e$. The projector $\widehat{P}_{\mathtt{h}}$ becomes the identity once pulled back to the generalized tangent space of $N$,
\begin{equation}
  (\pi_* + \sigma^*) \widehat{P}_{\mathtt{h}} (\pi_* + \sigma^*) = \id\,.
\end{equation}
Hence, \eqref{eqn:projectedgenLie} becomes eventually the Dorfman bracket on $T N \oplus T^* N$.

Let us now come back to the constraint \eqref{eqn:restrBwzw}. $\BW$ is a two-form on the coset $\mathcal{L}\backslash \PS$. We will decompose it into a horizontal part $\BW{}_{\mathtt{h}}$ and a vertical contribution in the directions of the Killing vectors $x_\alpha$. While $\varphi_\alpha$ is already a differential form on $\mathcal{L}\backslash\PS$, $A^\alpha$ has to be restricted accordingly. Its restriction is denoted as $a^{\alpha}$ and reads
\begin{equation}
  a^\alpha = \langle T^\alpha , m^{-1} \dd m \rangle\,.
\end{equation}
It is not hard to see that it is dual to the killing vector $x_\alpha$, because	$\iota_{x_\alpha} a^\beta = \delta_\alpha^\beta$ holds. In terms of this new quantity the of $\BW$ reads
\begin{equation}
	\BW = \BW{}_{\mathtt{h}} - 2 \varphi_\alpha \wedge a^\alpha\,.
\end{equation}
This relation implies a similar decomposition for $\HW$,
\begin{equation}
	\HW = \HW{}_{\mathtt{h}} - 2 \dd \varphi_\alpha \wedge a^\alpha - f_{\beta\gamma}{}^\alpha \varphi_\alpha \wedge a^\beta \wedge a^\gamma
\end{equation}
with
\begin{equation}
	\HW{}_{\mathtt{h}} = \dd \BW{}_{\mathtt{h}} + 2 \varphi_\alpha \wedge f^\alpha
		\quad \text{and} \quad
	f^\alpha = \dd a^\alpha + \frac12 f_{\beta\gamma}{}^\alpha a^\beta \wedge a^\gamma\,.
\end{equation}
While $\HW$ is closed by definition, this property does not carry over to its horizontal part. Instead we obtain
\begin{equation}\label{eqn:dHbar}
	\dd \HW{}_{\mathtt{h}} = 2 \left( \dd \varphi_\alpha + f_{\alpha\beta}{}^\gamma a^\beta \wedge \varphi_\gamma \right) \wedge f^\alpha \,. 
\end{equation}
It is instructive to rewrite this result in terms of doubled quantities. To this end, we introduce 
\begin{equation}
  \mathcal{A}^{\underline{A}} = \begin{pmatrix} 2 \varphi_\alpha & a^\alpha \end{pmatrix}
\end{equation}
and the corresponding field strength
\begin{equation}
  \mathcal{F}^{\underline{A}} = \dd \mathcal{A}^{\underline{A}} + \frac12 \mathcal{A}^{\underline{B}} \wedge \mathcal{A}^{\underline{C}} \mathbb{F}_{\underline{B}\underline{C}}{}^{\underline{A}}\,.
\end{equation}
The structure coefficients $\mathbb{F}_{\underline{A} \underline{B} \underline{C}}$ are totally antisymmetric. They are just constructed from $f_{\alpha\beta}{}^\gamma$, thus they generate a semi-abelian Drinfeld double. Hence, $\mathbb{F}$ is not just the restriction of the full structure coefficients $F$ of $\PS$ to the vertical generators. Using these new quantities \eqref{eqn:dHbar} can be written as
\begin{equation}\label{eqn:topoconstr}
  \dd \HW{}_{\mathtt{h}} = \mathcal{F}_a \wedge \mathcal{F}^a = \frac12 \langle \mathcal{F}, \mathcal{F} \rangle \,.
\end{equation}
This topological constraint appears also in the reduction of exact Courant algebroids on a principal bundle \cite{Baraglia:2013wua,Severa:2018pag}.

For heterotic string theory in flat spacetime, we find the similar expression
\begin{equation}\label{eqn:NS5het}
  \dd H = \frac{1}{30} \Tr F \wedge F
\end{equation}
where $F$ denotes the field strength corresponding to the SO(32) or E$_8 \times$E$_8$ gauge symmetry of the heterotic string. If we just consider a SU(2) subgroup and further restrict the field strength $F$ to a four-dimensional, euclidean subspace of the ten-dimensional Minkowski spacetime, we can construct instanton solutions for $F$. In this case the right hand side of \eqref{eqn:NS5het} does not vanish. After taking into account the gravitational backreaction a NS5-brane solution arises  in supergravity\cite{Strominger:1990et}. It would be interesting to see if NS5-branes arise in a similar fashion from \eqref{eqn:topoconstr}.

\section{Deformation theory}
\label{sec:Deformation}
In section~\ref{sec:isometries} we highlighted the close connection between para-Hermitian structures and Poisson-Lie symmetry. But how does T-duality arise in this picture? It is implemented by fixing $\eta$ and $\omega$ on $\PS$ in different ways. A natural question which arises in this context is how many dual backgrounds one can find for a given Lie group $\PS$. Equivalently, one might ask how many consistent choices there are for $\eta$ and $\omega$? This question can be approached by looking at small deformations. For all the examples discussed in this paper, these deformations can be studied at the level of the Lie algebra.

Assume that $K_0$ describes the action of $K$ at the identity of the Lie group $\PS$. It is an involutive (squares to one)  homomorphism of the Lie algebra $\mathfrak{p}$. We want to study small deformations of this map, denoted as $\delta K_0$. The first order deformation gives rise to the linear constraint
\begin{equation}\label{eqn:defconstr1}
  \delta K_0 K_0 + K_0 \delta K_0  = 0\,,
\end{equation}
since $K = K_0 + \delta K_0$ should still be an involution. This is sufficient for obtaining an almost para-Hermitian structure. But we also require the distribution $\Lt$ to be integrable. Give that $K_0$ is $\Lt$-integrable ($P_0[\Pt_0 X, \Pt_0 Y] = 0$), we find
\begin{equation}\label{eqn:defconstr2}
  \delta K_0 [ \Pt_0 T_A, \Pt_0 T_B ] - P_0 [ \delta K_0 T_A, \Pt_0 T_B ] - P_0 [ \Pt_0 T_A, \delta K_0 T_B ] = 0 \quad \forall \, T_A,\, T_B \in \mathfrak{p}\,,
\end{equation}
where $P_0 = \frac12(\id + K_0)$, $\Pt_0 = \frac12(\id - K_0)$ and $\brac$ denotes the Lie bracket of $\mathfrak{p}$ which is kept undeformed. 

As $\eta$ enters the definition of $K$, it is also deformed and we find that the deformation $\delta K_0$ is directly related to the deformations $\delta \eta$ and $\delta \omega$ of $\eta$ and $\omega$, respectively. The precise relation is captured by
\begin{equation}\label{eqn:deltaK}
  \delta K_0 = \eta^{-1}_0 ( \delta \omega_0 - \delta \eta_0 K_0 )\,.
\end{equation}
Furthermore, $\delta \eta_0$ has to be invariant under the adjoint action of Lie algebra $\mathfrak{p}$ which imposes additional constraints on it. It is instructive to choose a basis in which $K_0$ is diagonal,
\begin{equation}
  K_0{}^A{}_B = \begin{pmatrix} - \delta_a^b & 0 \\
    0 & \delta^a_b \end{pmatrix} \,.
\end{equation}
Here we are using $T_A = \begin{pmatrix} T^a & T_a \end{pmatrix}$ and $T^A = \begin{pmatrix} T_a & T^a \end{pmatrix}$ as an explicit basis for $\mathfrak{p}$. In this basis, \eqref{eqn:defconstr1} restricts $\delta K_0$ to be of the form
\begin{equation}
  \delta K_0{}^A{}_B = \begin{pmatrix} 0 & \delta K_0{}_{ab} \\
    \delta K_0{}^{ab} & 0 \end{pmatrix}
\end{equation}
and according to \eqref{eqn:deltaK} the two non-vanishing contributions are equivalent to
\begin{equation}
  \delta K_0{}_{ab} = \delta \omega_0{}_{ab} - \delta \eta_0{}_{ab}
    \quad \text{and} \quad
  \delta K_0{}^{ab} = \delta \omega_0{}^{ab} + \delta \eta_0{}^{ab} \,.
\end{equation}
The integrability condition \eqref{eqn:defconstr2} just constrains $\delta K_0{}^{ab}$ but not $\delta K_0{}_{ab}$. It gives rise to
\begin{equation}\label{eqn:tosolvefordefs}
  3 F^{[ab}{}_d \delta \omega_0{}^{c]d} - F^{bc}{}_d \delta \eta_0{}^{da} = 0 \,,
\end{equation}
where $F^{ab}{}_c$ denotes the structure coefficients of the maximally isotropic Lie subalgebra $\mathfrak{l}$. This equation has the trivial solution
\begin{equation}\label{eqn:trivialsol}
  \delta \omega_0{}^{ab} = F^{ab}{}_c \xi^c \quad \text{and} \quad \delta \eta_0{}^{ab} = 0\,,
\end{equation}
where $\xi^c$ is an arbitrary constant vector. They correspond to the adjoint $\mathfrak{l}$ action which is just a coordinate change for the unphysical directions. Thus, in finding solutions to \eqref{eqn:tosolvefordefs} we can ignore them. It is interesting to note that solving \eqref{eqn:tosolvefordefs} for $\delta \eta_0$=$0$ is equivalent to study the Lie algebra cohomology $H^2(\mathfrak{l},\mathbb{R})$. This observation is reassuring because cohomology problems are closely tied to deformation theory. Furthermore, the problem of identifying different T-dual backgrounds in DFT is also captured by Lie algebra cohomology \cite{Hassler:2016srl}. T-dual target spaces only arise from deformations with $\delta\eta_0$=$0$. This is because they are captured by O($D$,$D$) transformations which by definition leave $\eta$ invariant. In this case and when $\delta\omega_0^{ab}$ is invertable, the deformation is governed by the classical Yang-Baxter equation (CYBE) \cite{Borsato:2016pas}.

An interesting example, which is also relevant for the discussion in the next section, is SL(2,$\mathbb{C}$). It is a Drinfeld double that admits the decomposition into the two maximally isotropic subgroups SU(2) and B$_2$ (the Borel subgroup). Its ad-invariant pairing reads
\begin{equation}
  \eta_{AB} = \langle T_A, T_B \rangle = \mathrm{Im} \Tr ( T_A  T_B )\,,
\end{equation}
where $T_A$ are six complex, traceless 2$\times$2 matrices which generate SL(2,$\mathbb{C}$). This pairing permits two continuous deformations: either we scale $\eta_{AB}$ by a constant $C$ (this option always exists) or we rotate the trace by a complex phase $\rho$. In both cases we find that the $\delta\eta_0$ part of \eqref{eqn:tosolvefordefs} vanishes. For B$_2$ all solutions to this equation are of the form \eqref{eqn:trivialsol}. Thus, there are no non-trivial deformations for $\omega$.

\section{Examples}\label{sec:examples}
Let us now discuss a family of integrable deformations of the principal chiral model as an explicit example of the results we obtained above. Our discussion holds for a larger class of $\sigma$-models, but this particular subclass has a lot of additional structure and applications. We focus on the Lie group $\PS=$ SL(2,$\mathbb{C}$) which we already discussed in the last section. As we figured out there, its Lie algebra admits a two-parameter family of pairings
\begin{equation}
  \label{eqn:pairingSL2C}
  \eta_{AB} = C \, \mathrm{Im} \Tr ( e^{i \rho} T_A T_B )\,.
\end{equation}
Following \cite{Klimcik:2017ken}, we choose a basis for $\mathfrak{sl}(2,\mathbb{C})$ such that the structure in \eqref{eqn:eomPCM} of an integrable $\sigma$-model is manifest. To this end, the generators $T_A$ with $A=1,\dots,6$ are decomposed into $\mathcal{R}_a$ and $\mathcal{J}_a$ where $a$ runs from 1 to 3. On the other hand SL(2,$\mathbb{C}$) is the complexification of SU(2). Thus, one is inclined to choose the Lie algebra generators in terms of SU(2) generators, which we will denote as $t_a$. These two different bases are related by
\begin{equation}
  \begin{aligned}
    t_a &= \frac{\cos\rho \mathcal{R}_a + \sin\rho \sinh p \mathcal{J}_a}{2 \cosh p ( \cosh p \cos \rho + \sinh p )} \\
    i t_a &= i \frac{ (\cosh p + \cos \rho \sinh p ) \mathcal{J}_a - \sin \rho \mathcal{R}_a }{2 \cosh p ( \cosh p \cos \rho + \sinh p )}\,.
  \end{aligned}
\end{equation}
The pairing \eqref{eqn:pairingSL2C} for our particular choice of generators becomes
\begin{equation}\label{eqn:pairingSL2C2}
  \begin{aligned}
    \langle t_a, t_b \rangle &= C \sin\rho ( t_a, t_b ) &
    \langle i t_a, t_b \rangle &= C \cos\rho ( t_a, t_b ) &
    \langle i t_a, i t_b \rangle &= - C \sin\rho ( t_a, t_b )\,,
  \end{aligned}
\end{equation}
where $(t_a, t_b )$=$\Tr(t_a t_b)$ denotes the Killing form of SU(2). $\mathcal{E}$ has a particular simple form in the basis formed by $\mathcal{R}_a$ and $\mathcal{J}_a$, it just swaps them,
\begin{equation}\label{eqn:EYBWZ}
  \mathcal{R}^a = \mathcal{E} \mathcal{J}^a \quad \text{and} \quad
  \mathcal{J}^a = \mathcal{E} \mathcal{R}^a\,.
\end{equation}
This is all the information we need to completely pin down the $\mathcal{E}$-model. We conclude that there are three independent parameter: $C$ and $\rho$ in the pairing \eqref{eqn:pairingSL2C2} and $p$ which indirectly affects the generalized metric through \eqref{eqn:EYBWZ}. In the limit $\rho\rightarrow 0$ and $p\rightarrow \infty$, the SU(2) principal chiral model (PCM) arises. Its target space is a three-sphere whose radius is proportional to $\sqrt{C}$. There is no $H$-flux and we obtain a Born geometry. For any other values of $\rho$ and $p$, we obtain a two-parameter integrable deformation of the PCM which was dubbed Yang-Baxter Wess-Zumino model (YB WZ) \cite{Kawaguchi:2011mz,Kawaguchi:2013gma,Delduc:2014uaa,Klimcik:2017ken,Demulder:2017zhz}.

To identify a maximally isotropic subalgebra in $\mathfrak{p}$=$\mathfrak{g}^{\mathbb{C}}$, we employ the two Lie algebra homomorphisms
\begin{equation}
  R_\pm : \mathfrak{g} \rightarrow \mathfrak{p} \quad \text{with} \quad
  R_\pm = R \pm i\,,
\end{equation}
where $R$ is a map $R: \mathfrak{g} \rightarrow \mathfrak{g}$ called the R-matrix. As a homomorphism of a Lie algebra, $R_\pm$ has to satisfy $R_\pm [ x, y ]'$ = $[ R_\pm x, R_\pm y ]$ which gives rise to the modified classical Yang-Baxter equation (mCYBE)
\begin{equation}\label{eqn:mCYBE}
  [ R x, R y ] = R ( [ R x, y ] +  [x, R y] ) + [ x, y ] \qquad \forall \, x,\, y \, \in \mathfrak{g}\,.
\end{equation}
for the $R$-matrix. $\brac'$ is a deformed Lie bracket acting on the generators of $G$. Applying the mCYBE, we find that it is related to the original bracket by
\begin{equation}
  [ x, y ]' = [ R x, y ] + [ x, R y ] \,.
\end{equation}
Furthermore, the R-matrix is skew-symmetric
\begin{equation}
  ( R x, y ) = - ( x, R y )
\end{equation}
with respect to the Killing form of $\mathfrak{g}$. With these two properties, it is not hard to see that a maximally isotropic subgroup of SL(2,$\mathbb{C})$ is generated by
\begin{equation}\label{eqn:maxisofull}
  \tilde T_a = ( R_- - \tan \frac{\rho}{2} R_- R_+ ) t_a \,.
\end{equation}
This is all data we need in order to construct $\eta$, $\omega$, $\mathcal{H}$ and $\widehat{E}$ introduced in the previous sections. In order to present explicit expressions, all what remains to do is to choose a parameterization for the group element $\ell\in\mathcal{L}$ and a coset representative $m \in \mathcal{L}\backslash\PS$.

For the most general case with all three parameters of the deformation being non-zero, the resulting expressions are quite lengthy. Thus, we rather discuss the special case of $\rho=0$ in more detail. There the generators of $\mathcal{L}$ in \eqref{eqn:maxisofull} simplify to
\begin{equation}
  \tilde T_a = R_- t_a\,.
\end{equation}
Additionally, the generators $T_a := t_a$ form a maximally isotropic subalgebra because $\langle t_a, t_b \rangle$ in \eqref{eqn:pairingSL2C2} vanishes for $\rho=0$. For our examples $\mathfrak{g} = \mathfrak{su}(2)$ is relevant. We identify its generators $t_a = i \sigma_a$ with the Pauli matrices $\sigma_a$. The corresponding R-matrix is defined by
\begin{align}
  R t_1 &= t_2\,, &
  R t_2 &= -t_1\,, &
  R t_3 &= 0
\end{align}
and one can easily check that it indeed solves the mCYBE \eqref{eqn:mCYBE}. Now we are able to explicitly obtain the generators for the maximally isotropic subgroup $\mathcal{L}$. They read
\begin{align}
  \tilde T_1 &= \sigma_1 + i \sigma_2\,, &
  \tilde T_2 &= \sigma_2 - i \sigma_1\,, &
  \tilde T_3 &= - \sigma_3
\end{align}
and generate the Borel subgroup B$_2$ of SL(2,$\mathbb{C}$). We will show in the next subsection that this choice for $\mathcal{L}$ gives rise to the $\eta$-deformation three-sphere. Another choice for a maximal isotropic subgroup $\mathcal{L}$ is SU(2). It gives rise to the $\lambda^*$-deformation which is connected to the $\lambda$-deformation by an analytic continuation. We have a closer look at these two backgrounds in section~\ref{sec:lambdadef}. Finally, we come back the $\eta$-deformation of $S^3$ and apply the dressing coset construction to obtain the same deformation of $S^2$.

\subsection{\texorpdfstring{$\eta$}{Eta}-deformed three-sphere}\label{sec:etaS3}
We now explicitly present the construction of $\eta$ and $\omega$ for the maximally isotropic subgroup $\mathcal{L}$=B$_2$ of SL(2,$\mathbb{C}$). It is embedded in $\PS$=SL(2,$\mathbb{C})$ such that $\mathcal{L}\backslash\PS$=SU(2). A convenient parameterization for the group elements is
\begin{equation}\label{eqn:SU2andB2param}
  m = \frac1{\sqrt{2}} \begin{pmatrix}
    e^{i ( \phi_1 + \phi_2 )} \sqrt{1 + r} & e^{i (\phi_1 - \phi_2)} \sqrt{1 - r} \\
    -e^{- i ( \phi_1 - \phi_2 )} \sqrt{1 - r} &  e^{-i (\phi_1 + \phi_2)} \sqrt{1 + r}
  \end{pmatrix} \in \mathrm{SU}(2) \quad \text{and} \quad
  \ell = \begin{pmatrix}
      \tilde\xi & e^{i \tilde\phi} \tilde\varrho \\
      0 & 1/\tilde\xi
  \end{pmatrix}\in \mathrm{B}_2\,.
\end{equation}
The former describes the physical target space, a three-sphere. Since $\mathcal{L}\backslash \PS$ is a Lie group, $\HW$ vanishes and it is straightforward to calculate
\begin{equation}
  \begin{aligned}
    \eta = &\frac{C}{\sqrt{1-r^2} \tilde\xi^2} \Bigl( - \tilde\xi\sin\Delta\phi \,\dd r \dd\tilde\varrho + \tilde\xi\tilde\varrho\cos\Delta\phi \,\dd r \dd\tilde\phi - \tilde\varrho\sin\Delta\phi \,\dd r \dd\tilde\xi + \\
      & 2 (r^2 - 1) \, \dd \phi_2 ( \tilde\xi \cos\Delta\phi \,\dd\tilde\varrho + \tilde\xi \tilde\varrho \sin\Delta\phi \,\dd\tilde\phi + \tilde\varrho\cos\Delta\phi\dd\tilde\xi )  + 4 \sqrt{1 - r^2} \tilde\xi \,\dd\tilde\xi ( \dd\phi_1 + r \dd\phi_2 ) \Bigr)
  \end{aligned}
\end{equation}
and
\begin{equation}
  \begin{aligned}
    \omega = &\frac{C}{2 \sqrt{1-r^2} \tilde\xi^2} \Bigl( - \tilde\xi\sin\Delta\phi \,\dd r \wedge \dd\tilde\varrho + \tilde\xi\tilde\varrho\cos\Delta\phi \,\dd r \wedge \dd\tilde\phi - \tilde\varrho\sin\Delta\phi \,\dd r \wedge \dd\tilde\xi + \\
    & 2 (r^2 - 1) \, \dd \phi_2 \wedge ( \tilde\xi \cos\Delta\phi \,\dd\tilde\varrho + \tilde\xi \tilde\varrho \sin\Delta\phi \,\dd\tilde\phi + \tilde\varrho\cos\Delta\phi\dd\tilde\xi )  - 4 \sqrt{1 - r^2} \tilde\xi \,\dd\tilde\xi \wedge ( \dd\phi_1 + r \dd\phi_2 ) \Bigr)\,.
  \end{aligned}
\end{equation}
As we have two maximally isotropic subgroups, $\eta$ and $\omega$ give rise to a para-Hermitian structure.

In order to also calculate the generalized metric, we first have to relate the $\mathfrak{sl}(2,\mathbb{C})$ basis $\tilde T_a$ and $T_a := t_a$ to $\mathcal{R}_a$ and $\mathcal{J}_a$ for which we know the action of $\mathcal{E}$. A short calculation gives rise to
\begin{equation}
  \begin{aligned}
    T_a &=  \frac{\mathcal{R}_a}{2 e^p \cosh p} \qquad & 
    \mathcal{R}_a &= 2 e^p \cosh p T_a \\
    \tilde T_a &= \frac{R \mathcal{R}_a}{2 \cosh p e^p} - \frac{\mathcal{J}_a}{2 \cosh p} \qquad & 
    \mathcal{J}_a &= 2 \cosh p ( R T_a - \tilde T_a ) \,.
  \end{aligned}
\end{equation}
This allows us to calculate the contributions to the generalized metric
\begin{equation}
  \begin{aligned}
    \langle T_a , \mathcal{E} T_b \rangle &= - C e^{-p} (t_a, t_b) &
    \langle T_a , \mathcal{E} \tilde T_b \rangle &= - C e^{-p} (t_a, R t_b) \\
    \langle \tilde T_a , \mathcal{E} T_b \rangle &= \phantom{-} C e^{-p} (t_a, R t_b) \quad &
    \langle \tilde T_a , \mathcal{E} \tilde T_b \rangle &= - C \left( e^p (t_a, t_b) - e^{-p} ( t_a, R^2 t_b )  \right) \,.
  \end{aligned}
\end{equation}
We now could start to calculate the generalized frame field $\widehat{E}$ and eventually obtain the metric and the $B$-field on the target space $M=\mathcal{L}\backslash\PS$. But this was already done in full detail for \DFTwzw{} \cite{Demulder:2018lmj}. To see how our convention relates to the one used there, we raise the index on $\tilde T_a$ with the inverse $\kappa^{ab}$ of $\kappa_{ab}=\langle T_a, \tilde T_b\rangle=-C (t_a, t_b)$. We then obtain
\begin{equation}\label{eqn:compgenmetric}
  \begin{aligned}
    \langle T_a , \mathcal{E} T_b \rangle &= \phantom{-} e^{-p} \kappa_{ab} &
    \langle T_a , \mathcal{E} \tilde T^b \rangle &= e^{-p} \kappa_{ac} R^{cb} \\
    \langle \tilde T^a , \mathcal{E} T_b \rangle &=  - e^{-p} R^{ac} \kappa_{cb} \quad &
    \langle \tilde T^a , \mathcal{E} \tilde T^b \rangle &= e^p \kappa^{ab} - e^{-p} R^{ac} \kappa_{cd} R^{db}
  \end{aligned}
\end{equation}
with $R^{ab}$=$\kappa^{ac}\kappa^{bd}(t_c, R t_d)$, which is equivalent to (5.2) in \cite{Demulder:2018lmj} after identifying the deformation parameter $e^{-p}$=$\eta$. Thus, we conclude that the corresponding target space geometry is the $\eta$-deformation \cite{Klimcik:2002zj} of the SU(2) PCM. It has a non-trivial $H$-flux and therefore is not a Born geometry.

\subsection{\texorpdfstring{$\lambda^*$}{Lambda*}- and \texorpdfstring{$\lambda$}{lambda}-deformed three-sphere}\label{sec:lambdadef}
Another maximally isotropic subgroup of the pairing \eqref{eqn:pairingSL2C} for $\rho$=0 is SU(2). The canonical choice for a coset representative would be B$_2$, which we already used for $\mathcal{L}$ in the last subsection. Here however, we identify the coset rather with the hermitian 2$\times$2 matrices
\begin{equation}\label{eqn:mhermitian}
  m = \begin{pmatrix} e^{\phi_1} \sqrt{1 + r^2} & -i r e^{i \phi_2} \\
    i r e^{-i\phi_2} & e^{-\phi_1} \sqrt{1 + r^2}
  \end{pmatrix}\,.
\end{equation}
For $\mathcal{L}$ we use the SU(2) group element from \eqref{eqn:SU2andB2param} but with tilded coordinates. This particular splitting of SL(2,$\mathbb{C}$) result in a non-vanishing
\begin{equation}
  \HW =  4 C r \, \dd r \wedge \dd \phi_1 \wedge \dd \phi_2 \,.
\end{equation}
It can be written in terms of the $B$-field
\begin{equation}
  \BW = 2 C r^2 \, \dd \phi_1 \wedge \dd \phi_2
\end{equation}
as $\HW$=$\dd \BW$. Evaluating \eqref{eqn:etaXY} and \eqref{eqn:omegaXY} gives rise to
\begin{equation}
  \begin{aligned}
  \eta = 2 C &\sqrt{\frac{1+r^2}{1-\tilde r^2}} \Bigl( \frac{\cosh\phi_1 \dd r}{1+r^2} \bigl( 2 (1 - \tilde r^2)\sin\Delta\tilde\phi \, \dd\tilde \phi - \cos\Delta\tilde\phi \, \dd\tilde r  \bigr) + r \cos\Delta\tilde\phi \sinh\phi_1 \, \dd \phi_1 \dd\tilde r \\ 
    & - 2 r (1-\tilde r^2) \sin\Delta\tilde\phi \sinh\phi_1 \, \dd\phi_1 \dd\tilde\phi_1 + 2 \sqrt{(1+r^2)(1-\tilde r^2)} \, \dd\phi_1 ( \tilde r \, \dd\tilde \phi_1 + \dd \tilde \phi_2 ) \\
  & + r \cosh\phi_1 \sin\Delta\tilde\phi \, \dd\phi_2 \dd\tilde r + 2 r (1-\tilde r^2) \cos\Delta\tilde\phi \cosh\phi_1 \, \dd\phi_2 \dd\tilde\phi_1 \Bigr)
  \end{aligned}
\end{equation}
and
\begin{equation}
  \begin{aligned}
  \omega = C &\sqrt{\frac{1+r^2}{1-\tilde r^2}} \Bigl( \frac{\cosh\phi_1 \dd r}{1+r^2} \wedge \bigl( 2 (1 - \tilde r^2) \sin\Delta\tilde\phi \, \dd\tilde \phi - \cos\Delta\tilde\phi \, \dd\tilde r  \bigr) + r \cos\Delta\tilde\phi \sinh\phi_1 \, \dd \phi_1 \wedge \dd\tilde r \\ 
    & - 2 r (1-\tilde r^2) \sin\Delta\tilde\phi \sinh\phi_1 \, \dd\phi_1\wedge\dd\tilde\phi_1 + 2 \sqrt{(1+r^2)(1-\tilde r^2)} \, \dd\phi_1\wedge( \tilde r \, \dd\tilde \phi_1 + \dd \tilde \phi_2 ) \\
  & + r \cosh\phi_1 \sin\Delta\tilde\phi \, \dd\phi_2\wedge\dd\tilde r + 2 r (1-\tilde r^2) \cos\Delta\tilde\phi \cosh\phi_1 \, \dd\phi_2\wedge\dd\tilde\phi_1 \Bigr) + 2 C r^2 \,\dd\phi_1\wedge\dd\phi_2 \,.
  \end{aligned}
\end{equation}
As $\HW$ does not vanish, $\eta$ and $\omega$ only give rise to a $\Lt$-integrable structure.

The target space geometry which originates from choosing $\mathcal{L}$=SU(2) is Poisson-Lie T-dual to the $\eta$-deformation discussed in the last subsection. It is called $\lambda^*$-deformation because it is closely related to the $\lambda$-deformation \cite{Sfetsos:2013wia}. Instead of deforming a PCM, the latter starts from a WZW-model. An analytic continuation allows to transition from the $\lambda^*$- to the $\lambda$-deformation \cite{Hoare:2015gda,Sfetsos:2015nya,Klimcik:2015gba}. It takes $m$ from a hermitian matrix to a unitary one. For $m$ in \eqref{eqn:mhermitian} this analytic continuation reads
\begin{align}
  \phi_1 \rightarrow i \phi_1 \qquad &\text{and} \qquad r \rightarrow i r\,.
  \intertext{and results in}
  \eta \rightarrow - i \eta' \qquad &\text{and} \qquad \omega \rightarrow -i \omega'
\end{align}
where $\eta'$ and $\omega'$ describe the $\lambda$-deformation on $\PS$= SU(2)$\times$SU(2) with $\mathcal{L}$=SU(2)$_{\mathrm{diag}}$. The corresponding coset representative is $(m',m'{}^{-1})$ where $m'$ is a SU(2) element that squares to $m$ in \eqref{eqn:mhermitian} after the analytic continuation.

\subsection{\texorpdfstring{$\eta$}{eta}-deformed two-sphere}
Let us finally come to the dressing coset construction. Our starting point is the $\eta$-deformation discussed in section~\ref{sec:etaS3}. Its target space is the group manifolds SU(2) which is isomorphic to the three-sphere $S^3$. The latter admits a Hopf  fibration $S^1 \hookrightarrow S^3 \rightarrow S^2$. Using the parameterization \eqref{eqn:SU2andB2param}, we find that the coordinates $r$ and $\phi$ describe the base $S^2$ which can be embedded into $\mathbb{R}^3$ with the Cartesian coordinates $y^1$, $y^2$ and $y^3$ according to
\begin{align}
  y^1 &= r &
  y^2 &= \sqrt{1-r^2} \cos(2 \phi_1) &
  y^3 &= \sqrt{1-r^2} \sin(2 \phi_1)\,.
\end{align}
This embedding shows that $r\in[-1,1]$ and $\phi_1 \in[0,\pi)$. Furthermore, we notice that the right-action of an U(1) element $f=e^{T_3 \Delta \phi_2}$ just shifts the fiber coordinate $\phi_2$ by $\Delta \phi_2$ but does not affect the base. Thus, we identify the $S^2$ with the coset SU(2)/U(1). Since the U(1) we mod out is an isotropic subgroup of SL(2,$\mathbb{C}$) with the pairing \eqref{eqn:pairingSL2C} ($\rho=0$), we have satisfied all requirements for the dressing coset construction.

We have seen in the last two subsection that the explicit expressions for $\eta$ and $\omega$ are lengthy. The same holds for their projection to the horizontal subspace $\eta_{\mathtt{h}}$ and $\omega_{\mathtt{h}}$. Hence, we will not present them here. However, we have checked that they satisfy \eqref{eqn:Khorizontal}. Instead we rather discuss the generalized frame field $\iota_{\bar A} \widehat{E}^{-1}_{\mathtt{h}}$ in more detail. The index $\bar A$ captures the generators of the horizontal subspace $T_{\bar A} = \begin{pmatrix} \tilde T^1 & \tilde T^2 & T_1 & T_2 \end{pmatrix}$. Applying \eqref{eqn:Ehatinv}, we eventually find
\begin{equation}
  \begin{aligned}
    \iota_1 \widehat{E}^{-1}_{\mathtt{h}} &= \sqrt{1-r^2} \left( \phantom{-} 2 \sin(2\phi_2) \partial_r +
      \frac{\cos(2\phi_2)}{1-r^2} ( \partial_{\phi_1} - r \partial_{\phi_2}) \right)\\
    \iota_2 \widehat{E}^{-1}_{\mathtt{h}} &= \sqrt{1-r^2} \left( - 2 \cos(2\phi_2) \partial_r + 
      \frac{\sin(2\phi_2)}{1-r^2}( \partial_{\phi_1} - r \partial_{\phi_2} ) \right)\\
    \iota^1 \widehat{E}^{-1}_{\mathtt{h}} &= \sqrt{1-r^2} \left( - \frac{(1-r)\cos{2\phi_2}}{C} \partial_r +
    \frac{\sin(2\phi_2)}{2 C(1 + r)} (\partial_{\phi_1} + \partial_{\phi_2} ) + \frac{\sin(2\phi_2)}{2(1-r^2)} \dd r + \cos(2\phi_2) \dd \phi_1 \right)\\
    \iota^2 \widehat{E}^{-1}_{\mathtt{h}} &= \sqrt{1-r^2} \left( - \frac{(1-r) \sin{2\phi_2}}{C} \partial_r -
      \frac{\cos(2\phi_2)}{2 C (1+r)} ( \partial_{\phi_1} + \partial_{\phi_2} ) - \frac{\cos(2\phi_2)}{2( 1-r^2)} \dd r + \sin(2\phi_2) \dd \phi_1 \right)\,.
  \end{aligned}
\end{equation}
This generalized frame field satisfies the frame algebra \eqref{eqn:framealgebraDC} where the restricted structure coefficients $F_{\bar A\bar B\bar C}$ vanish. We can now use this generalized frame field to obtain the generalized metric $\widehat{\mathcal{H}}$ on $T N \oplus T^* N$ with N = SU(2)/U(1), namely
\begin{equation}
  \widehat{\mathcal{H}} \left( (\pi_* + \sigma^*) \iota_{\bar A} \widehat{E}^{-1}_{\mathtt{h}}, 
  (\pi_* + \sigma^*) \iota_{\bar B} \widehat{E}^{-1}_{\mathtt{h}} \right) = \langle T_{\bar A} , \mathcal{E} T_{\bar B} \rangle\,.
\end{equation}
Since we have chosen an adapted coordinate system, the pushforward map $\pi_*$ and the pullback map $\sigma^*$ are simple. They just chop off the $\partial_{\phi_2}$ and $\dd \phi_2$ contributions. Furthermore, we have to set $\phi_2$ to a fixed value, for example $\phi_2$=0. Because $\mathcal{E}$ is invariant under the U(1) we gauge, this choice does not affect $\widehat{\mathcal{H}}$. The right hand side of the equation was already calculated in \eqref{eqn:compgenmetric}. Finally, we read of the metric
\begin{equation}
  \dd s^2 = \frac{C}{e^{p}(1 + e^{-2 p} r^2)} \left( \frac{\dd r}{2(1-r^2)} + 2 (1-r^2) \dd \phi_1 \right)
\end{equation}
and the $B$-field
\begin{equation}
   B = \frac{C r}{2 e^{2 p}(1 + e^{-2 p} r^2)} \dd \phi_1 \wedge \dd r \,.
\end{equation}
from the generalized metric in \eqref{eqn:genmetricGB}. The resulting expressions match nicely with \cite{Hoare:2017ukq} once we remember that the deformation parameter $\eta=e^{-p}$.

\section{Conclusions}
In this article we have shown that all group manifolds which appear in the construction of Poisson-Lie $\sigma$-models give rise to an almost para-Hermitian structure. It is formed by an even-dimensional Lie group $\PS$ endowed with a para-complex structure $K$ and a pseudo-Riemannian metric $\eta$ of split signature. Together they give rise to an almost symplectic form $\omega$. Only in special cases, this structure is completely integrable and this paper shows that the notion of half-integrability is sufficient to capture all features of the Poisson-Lie $\sigma$-model. Because half-integrable structures are central in our discussion, let us quickly recap how they arise. The tangent bundle $T\PS$ of a para-Hermitian manifold is naturally split into two subdistributions $L$ and $\Lt$. Unlike in the Hermitian case, they can be independently integrable. Hence, it is possible to have only one integrable distribution which is then call $L$- or $\Lt$-integrable. Recent works \cite{Freidel:2017yuv,Svoboda:2018rci,Freidel:2018tkj} focused on setups where at least $L$ has to be integrable. However, we find that on group manifolds (and there is no obvious reason why this results should not also hold for more general cases) instead of $L$-integrability, $\Lt$-integrability is the only requirement to make contact with the $\sigma$-model. $L$-integrability breaks down once the $\sigma$-model admits a WZ-term.

The Lie group $\PS$ can be view from two different perspectives. For the closed string worldsheet theory it plays the role of the phase space, as demonstrated in section~\ref{sec:Dstructure2}. But it also acts as doubled target space manifold when approached from DFT. We show that it is equipped with a generalized differentiable structure, the so called D-structure. Most importantly for applications in physics, this structure includes a D-bracket whose extension to higher rank tensors gives rise to the generalized Lie derivative. It captures the global symmetries on the worldsheet theory and the local symmetries of the target space, namely diffeomorphisms and $B$-field transformations. In the case where the $H$-flux on the target space vanishes, the generalized metric $\mathcal{H}$ together with $\eta$ and $\omega$ from the almost para-Hermitian structure form a Born geometry.

A para-Hermitian or Born manifold is the target space of a doubled $\sigma$-model. In the case of group manifolds this model is the so called $\mathcal{E}$-model \cite{Klimcik:1995dy,Klimcik:1996nq,Klimcik:2015gba}. It incorporates a WZ term which is governed by the three form $F$ on $\PS$. The latter encodes the structure constants of the Lie algebra $\mathfrak{p}$ of $\PS$ and measures the obstruction of $\omega$ to close, $F=\dd\omega$. From a string theory point of view it implements the various fluxes on the target space. More specifically, we established the precise relations between para-Hermitian and Born geometry, generalized geometry, \DFTwzw{} and the $\mathcal{E}$-model. Furthermore, we showed that Poisson-Lie symmetry is related to the isometry group of the generalized metric $\HH$ and the integrability of the $\sigma$-model is discussed.

Another interesting setup that we investigated is that of a dressing coset\cite{Klimcik:1996np,Squellari:2011dg,Klimcik:2019kkf} which arises when considering gauged Poisson-Lie $\sigma$-models. Here an isotropic subgroup of $\PS$ is gauged and the gauge connection provides a splitting of the doubled target space into a horizontal and vertical subspace. The vertical projections of tensors just captures the gauge symmetry of the model. On the horizontal subspace we construct an almost para-Hermitian structure and, in case of vanishing $H$-flux, a Born geometry as before. We also consider deformations which remain half-integrable. They are a powerful tool to explore the moduli space of T-dual backgrounds, because T-dual backgrounds arise if $\PS$ admits different $\Lt$-integrable structures for the same $\eta$.

Besides the conceptual insights, the most practical outcome of our work is that we provide an explicit construction of a huge class of para-Hermitian geometries. We also show how the corresponding generalized frame fields are constructed. Using them for generalized Scherk-Schwarz reductions \cite{Grana:2012rr,Geissbuhler:2011mx,Aldazabal:2011nj} results in a large class of new consistent truncations. Explicit examples we are presenting are integrable deformations of the three-sphere and two-sphere. They are obtained by considering various subgroups of $\PS$=SL(2,$\mathbb{C})$. These examples nicely display the structures and features discussed in this paper.

\acknowledgments
We thank Saskia Demulder and Daniel Thompson for comments on the draft and helpful discussions. FH would like to express a special thanks to the Mainz Institute of Theoretical Physics where this work was finished during the program ``Holography, Generalized Geometry and Duality''. FH is partially supported by the Spanish Government Research Grant FPA2015-63667-. The research of D.L. is supported by the Excellence Cluster ''ORIGINS: From the Origin of the Universe
to the First Building Blocks of Life''. FJR is supported by the Max-Planck-Society.

\appendix
\pagebreak
\bibliography{literatur}
\bibliographystyle{JHEP}
\end{document}